\documentclass[
preprint,
showpacs,amsmath,amssymb,aps,pra]{revtex4-1}

\usepackage{graphicx}
\usepackage{dcolumn}
\usepackage{bm}

\usepackage{color}
\usepackage[pdfpagelabels,pdftex,bookmarks,breaklinks]{hyperref}
\definecolor{darkblue}{RGB}{0,0,127} 
\definecolor{darkgreen}{RGB}{0,150,0}
\hypersetup{colorlinks, linkcolor=blue, citecolor=magenta, filecolor=red, urlcolor=blue}
\hypersetup{pdfpagemode=UseNone} 
\hypersetup{pdfstartview=FitH} 

\newcommand{\eref}[1]{(\ref{#1})}
\newcommand{\etal}{{\it et al.}}
\newcommand{\hop}[2]{\hat{#1}_{#2}}
\newcommand{\rb}{\bar{r}}

\begin{document}

\title{Entanglement enhancement through multi-rail noise reduction for continuous-variable
measurement-based quantum information processing}

\author{Yung-Chao Su and Shin-Tza Wu}
\email{phystw@gmail.com}
\affiliation{
Department of Physics, National Chung Cheng University, Chiayi 621, Taiwan}

\date{\today}

\begin{abstract}
We study theoretically the teleportation of controlled-phase (CZ)
gate through measurement-based quantum information processing for
continuous-variable systems. We examine the degree of entanglement
in the output modes of the teleported CZ-gate for two classes of
resource states: the canonical cluster states that are constructed
via direct implementations of two-mode squeezing operations, and the
linear-optical version of cluster states which are built from
linear-optical networks of beam splitters and phase shifters. In
order to reduce the excess noise arising from finite-squeezed
resource states, teleportation through resource states with
different multi-rail designs will be considered and the enhancement
of entanglement in the teleported CZ-gates will be analyzed. For
multi-rail cluster with an arbitrary number of rails, we obtain
analytical expressions for the entanglement in the output modes and
analyze in detail the results for both classes of resource states.
At the same time, we also show that for uniformly squeezed clusters
the multi-rail noise reduction can be optimized when the excess
noise is allocated uniformly to the rails. To facilitate the
analysis, we develop a trick with manipulations of quadrature
operators that can reveal rather efficiently the measurement
sequence and corrective operations needed for the measurement-based
gate teleportation, which will also be explained in detail.
\end{abstract}

\pacs{03.67.Lx, 42.50.Ex}

\maketitle

\section{\label{sec:int}Introduction}
Measurement has been the foundation of physical science since Galileo Galilei's time \cite{gali}.
It has also played a pivotal role in the development of quantum mechanics.
Studies of quantum measurements have had great
impacts on our understanding of the foundation of quantum mechanics \cite{WZ83}.
At the same time, manipulations of quantum systems through measurements have been
exploited in many aspects of modern quantum technologies \cite{BK92,WM10}.

In quantum information sciences, quantum information processing
typically involves sequence of transformations that are represented
by unitary operators acting on the information carriers \cite{NC00}.
Experimental realizations for these processes, however, have been
challenging due to the highly demanding control over coherence
\cite{Di00,Br01}. Measurement-based quantum information processing
offers a partial solution to such difficulties \cite{RB01}. The
quantum information processing in this approach is effected by
appropriately designed sequence of local measurements over highly
entangled resource states. By feeding forward the measurement
outcomes, gate operation over the input state can be achieved
through appropriate corrective operations upon the output. In this
way, the intended gate operation can be ``teleported" to the output
of the entangled resource state \cite{Ni06,Br09,note1}.

Originally, measurement-based quantum information processing was
proposed for discrete-variable systems, i.e., systems with
finite-dimensional state space \cite{RB01}. It was later extended to
systems with continuous degrees of freedom, which are often referred
to as ``continuous-variable" (CV) systems \cite{Zh06,Me06}. Since
then, optical systems have been providing an appealing platform for
CV quantum information processing through the measurement-based
scheme due to their advantages in not only state preparation, but
also in the detection and manipulation of states
\cite{BP03,BrvL05,Ce07,We12}. Nevertheless, the archetypal resource
state for measurement-based CV quantum information processing
requires two-mode squeezing operations over a set of quantum modes
prepared in the zero-momentum eigenstate to form a highly entangled
state commonly known as a ``cluster state" (or a ``graph state")
\cite{Me06} over which definite correlations among the ``nodes" of
the cluster are imposed by the entanglement. Since an ideal momentum
eigenstate demands infinite squeezing (hence infinite energy), only
approximated, finite-squeezed momentum states are available in the
laboratories. In practice, therefore, CV cluster states always have
non-ideal correlations among their nodes due to finite squeezing,
which in turn deteriorate the quality of the gate teleportation with
such resource states. Moreover, although two-mode squeezing
operation can be implemented through inline-squeezers and
beam-splitters \cite{Br05,note2}, the rapid growing demand for its
implementation along with the size of the canonical CV cluster
renders the scheme experimentally challenging for practical
applications. In order to alleviate the demand for inline squeezers,
van Loock and colleagues proposed in Ref.~\onlinecite{vL07} a
linear-optical approach to the construction of CV cluster states,
which consists of finding appropriate linear-optical network that
can implement the desired cluster correlations over a collection of
offline squeezed modes through combinations of beam-splitters and
phase shifters. Making use of this approach, experimental
demonstrations for measurement-based single-mode operations
\cite{Uk11a}, two-mode operation \cite{Uk11b}, and their sequential
operation \cite{Su13} have been achieved through linear-optical CV
cluster states. However, the linear-optical approach remains to
suffer from difficulties with scalability, since the number of
optical elements in the optical network increases rapidly with the
size of the cluster. Subsequent endeavors to improve the scalability
of optical CV cluster states via encoding in the frequency domain
\cite{Me08} or in the time domain \cite{Me10,Me11} have made a lot
of progress in recent years. It has also been shown that
fault-tolerant quantum computing in the measurement-based CV scheme
can be achieved with finite squeezing, although at a level that is
still experimentally demanding \cite{Me14}. These developments have
made measurement-based CV approach a promising direction for
realistic quantum computing.

\begin{figure}
\includegraphics*[width=90mm]{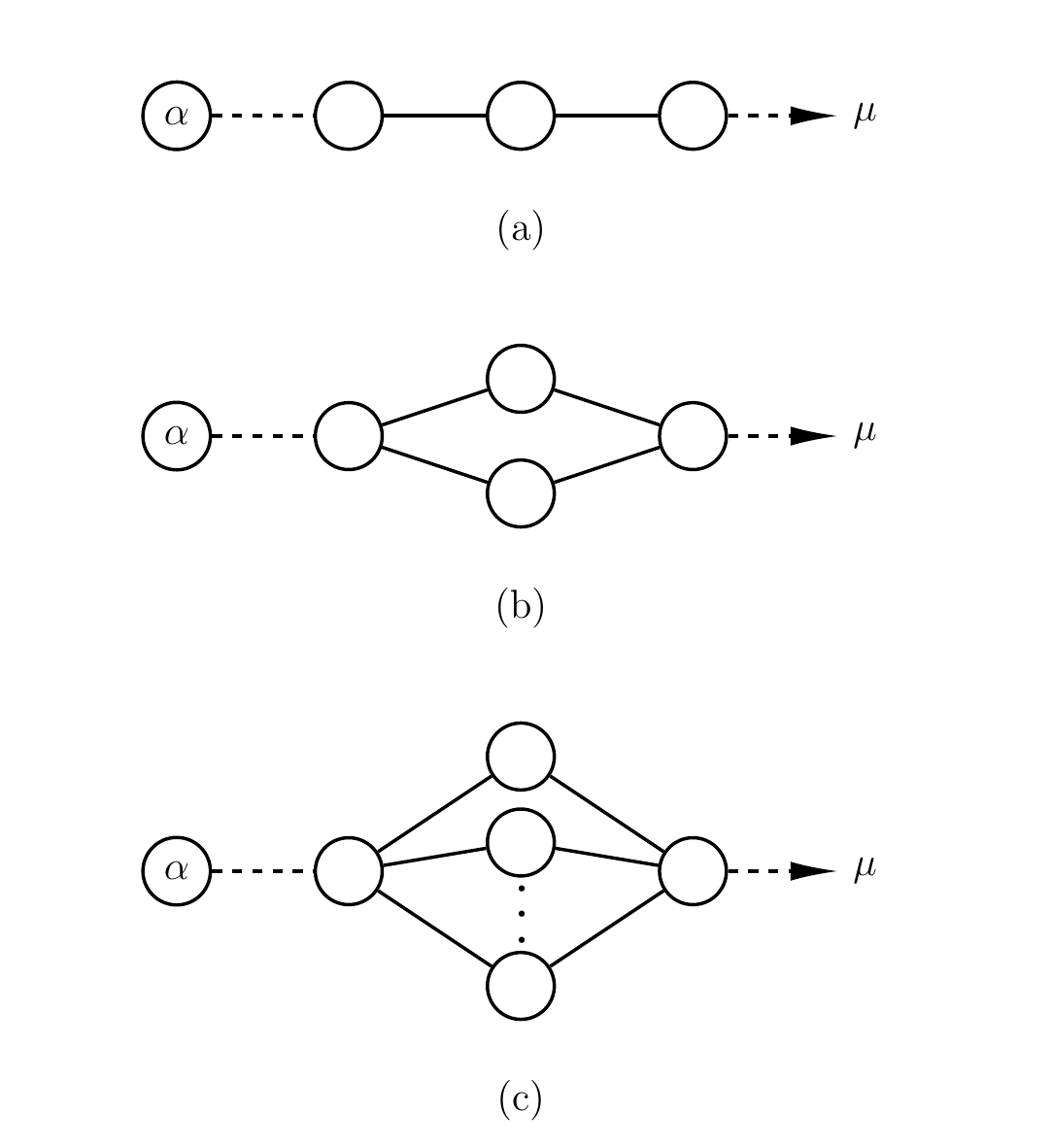}
\caption{Excess-noise reduction for measurement-based teleportation through multi-rail designs for CV cluster states
proposed by van Loock \etal~\cite{vL07}
with (a) single-rail, (b) two-rail, and (c) general multi-rail structures. Here each circle represents a qumode, with
the one labeled $\alpha$ the input mode and those unlabeled the cluster modes. The solid lines connecting the
cluster nodes represent two-mode correlations which can be established through two-mode squeezing operations or linear-optical
networks (see the text). In each figure, the coupling between the input mode $\alpha$ and the
cluster is represented by a dashed line. To teleport the state of $\alpha$ towards the right of the cluster, a sequence
of quadrature measurements must be applied to the cluster modes. Upon feeding forward the measurement outcomes
and applying the corresponding corrective operations to the rightmost node of the cluster, as indicated by the arrow, one would get the
output mode $\mu$ which holds the teleported state.
\label{fig:MR_tel}}
\end{figure}

As with canonical cluster states, the linear-optical cluster states (the qualifier
``CV" will henceforth be omitted when it is clear in context)
are also non-ideal in practice, since the initial offline squeezed modes for the cluster
are always squeezed finitely. This results
in imperfect cluster correlations among the cluster nodes, and cause ``excess noise" in
measurement-based teleportation through such resource states. As a remedy,
van Loock and coworkers proposed in Ref.~\onlinecite{vL07} a multi-rail design
(see Fig.~\ref{fig:MR_tel}) for the CV cluster state that is capable of reducing
excess noise in teleportation. They showed that, with
increasing number of rails in the design, the excess noise becomes progressively
smaller in one of the quadratures of the teleported mode. A comparison
for the noise reduction was drew in Ref.~\onlinecite{vL07} for single-mode
teleportation through multi-rail canonical cluster states and
linear-optical cluster states. Here we extend the consideration by examining the
teleportation of a two-mode entangling gate, the controlled-phase
(or controlled-$Z$, abbreviated CZ) gate,
through these two classes of CV cluster states. In particular, with the reduction
of excess noise through the multi-rail design, we expect to find improvements in the
quality of the teleported CZ-gate, which should reveal in the enhanced entanglement
in the output modes of the cluster. Since the multi-rail design can increase the size
of the cluster massively (see Fig.~\ref{fig:MR_tel}), it is necessary to have a way
to find out the teleportation scenario efficiently. To this end, we devise an
intuitive Heisenberg approach that can reveal the measurement
sequence and the corresponding corrective operations for the teleportation through
simple manipulations of the quadrature operators \cite{Su15}. We are able to find out
analytical results for the entanglement of the teleported CZ-gates for both canonical
and linear-optical cluster states with general multi-rail configurations.
In the process of this analysis, we will also show that noise reduction in the multi-rail scheme can be
optimized when the excess noise is distributed uniformly to each of the multi-rails.

We will begin in the following section by providing background
information for CV cluster states necessary to our calculations and
setting up the notations. In particular, we will explain the CZ-gate
teleportation in measurement-based CV quantum information processing
using a Heisenberg approach that involves certain tricks for
manipulating the quadrature operators which will be essential to our
calculations. We will then study in Sec.~\ref{sec:CZ} the CZ-gate
teleportation through canonical cluster states and linear-optical
cluster states in two different subsections, where the
noise-reduction mechanism in the multi-rail scheme will also be
analyzed. Comparison of the results for the two classes of resource
states will be presented at the later part of Sec.~\ref{sec:CZ}.
Finally, we close in Sec.~\ref{sec:concl} with brief comments on the
results and their possible extensions. For the sake of clarity of
our presentation, a number of details of our results have been
relegated to the Appendices.

\section{\label{sec:form} Formulation}
In CV systems, each quantum mode (or ``qumode", in analogy with ``qubit" for quantum bit) is
described as a quantized harmonic oscillator \cite{BP03}. For a qumode $k$, if the annihilation
operator is $\hop{a}{k}$, the corresponding quadrature amplitude (or ``position") and
quadrature phase (or ``momentum") operators are then given, respectively, by \cite{WM08}
\begin{eqnarray}
\hop{q}{k} \equiv \frac{\hop{a}{k}+\hop{a}{k}^\dagger}{2}
\quad\mbox{and}\quad
\hop{p}{k} \equiv  \frac{\hop{a}{k}-\hop{a}{k}^\dagger}{2i} \, ,
\label{qp_def}
\end{eqnarray}
where $\dagger$ indicates Hermitian conjugation. These quadrature
operators possess continuous eigenvalues and the corresponding
eigenstates provide bases for encoding CV quantum information
\cite{BP03}. For any two qumodes $k$ and $l$, it follows from
\eref{qp_def} and the commutation relation
$[\hop{a}{k},\hop{a}{l}^\dagger] = \delta_{kl}$ that
\begin{eqnarray}
[\hat{q}_k, \hat{p}_l]=\frac{i}{2} \delta_{kl} \, ,
\label{qp_cmm}
\end{eqnarray}
which corresponds to the canonical commutation relations for position and momentum operators
in mechanical systems with $\hbar=1/2$.

Canonical cluster states for CV systems are constructed by applying
CZ-gates over a set of momentum-squeezed vacuum modes that constitute the ``nodes" of the cluster.
Explicitly, the CZ-gate that acts on modes $k$ and $l$ is given by
\begin{eqnarray}
CZ_{kl} = e^{2i\,\hat{q}_k\hat{q}_l} \, ,
\label{cz}
\end{eqnarray}
which is clearly symmetrical between modes $k$ and $l$. By virtue of the commutation relation \eref{qp_cmm},
it follows that the CZ-gate operation on the quadrature operators for mode $l$ yields
\begin{eqnarray}
CZ_{kl}^\dag\,\hat{q}_l\,CZ_{kl} = \hat{q}_l \, ,
\qquad
CZ_{kl}^\dag\,\hat{p}_l\,CZ_{kl} = \hat{p}_l + \hat{q}_k \, ,
\label{cz_map}
\end{eqnarray}
and likewise for mode $k$. Therefore, we see that the CZ-gate has
established correlations between the quadrature operators of the two
modes. Experimentally, the transformation \eref{cz_map} corresponds
to a quantum non-demolition (QND) process, where the position
operator is unchanged, while the momentum operator picks up a shift
from the other mode \cite{WM08}. The CZ-gate is thus also often
referred to as a ``QND-gate" \cite{Zh06}. For definiteness,
hereafter we will refer to the operation that connects the nodes of
a canonical cluster the QND-gate, while the gate \eref{cz_map} to be
teleported using measurement-based schemes the CZ-gate, although the
two indeed function identically.

According to \eref{cz_map}, if a collection of modes
with annihilator operators $\hop{\bar{a}}{k}=\hop{\bar{q}}{k}+i\,\hop{\bar{p}}{k}$ are coupled through
QND-gates, one then has for the resultant mode $k$
\begin{eqnarray}
\hat{q}_k &=& \hat{\bar{q}}_k \, ,
\nonumber\\
\hat{p}_k &=& \hat{\bar{p}}_k + \sum_{l\in N_k} \hat{\bar{q}}_l \, ,
\label{qp_QND}
\end{eqnarray}
where $N_k$ denotes the set of modes that are coupled with mode $k$ through QND-gates. For initial modes that are
momentum-squeezed vacuum states, one can define
\begin{eqnarray}
\hat{\delta}_k &\equiv& \hat{p}_k - \sum_{l\in N_k} \hat{q}_l
\nonumber\\
&=& \hop{\bar{p}}{k} = e^{-r_k} \hat{p}^{(0)}_k \, ,
\label{noise_can}
\end{eqnarray}
where $r_k$ is the squeezing parameter for mode $k$, $\hat{p}^{(0)}_k$ is the momentum operator for the
respective vacuum mode, and we have used \eref{qp_QND} in arriving at the second line.
It is then clear that in the ideal, infinite-squeezing limit $r_k\rightarrow \infty$
for all $k$, the $\hop{\delta}{k}$'s would vanish identically, and the state would approach an
ideal cluster state, which has perfect quadrature correlations among its nodes \cite{Zh06,We12}.
These operators $\hop{\delta}{k}$ thus represent the noise in the cluster correlations
due to finite-squeezed initial modes, and are often called the ``excess-noise" operators, or the ``nullifiers"
of the cluster state, since the ideal cluster state is an eigenstate of these operators with eigenvalue zero \cite{We12}.
Depending on the context, we will use both terms for $\hop{\delta}{k}$ interchangeably and
at times, for brevity, also refer to them simply as the ``noise operators".

\begin{figure}
\includegraphics*[width=75mm]{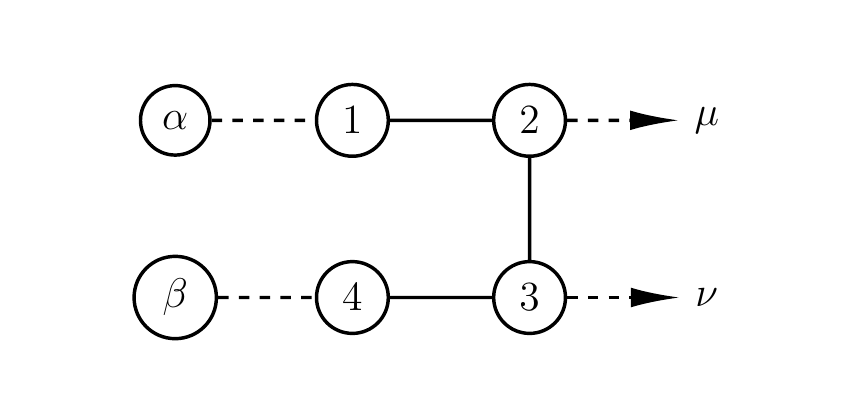}
\caption{Teleportation of a CZ-gate through a linear four-mode cluster. Here and in all figures
below, the input modes $\alpha$, $\beta$ are coupled to the cluster nodes via either
a QND-coupling or a beam-splitter coupling (see the text), which is represented by a dashed-line here.
To the right of the cluster, the output modes $\mu$ and $\nu$ are generated from nodes 2 and 3,
respectively, where the arrows indicate appropriate corrective operations
in accordance with measurement outcomes in the measurement sequence. For an ideal cluster state,
the teleported state registered by $\mu$, $\nu$ would be identical to that of the input modes
$\alpha$, $\beta$ operated with a CZ-gate. Depending on the context, the cluster
state here (and also in other figures of this paper) can be a canonical one or a linear-optical one.
\label{fig:L4-1}}
\end{figure}

For the teleportation of CZ-gate using different types of CV cluster states that we will discuss
in the following, rather than the Schr\"{o}dinger approach \cite{Me06}, we will resort to a
Heisenberg approach that keeps track of the teleportation
process through the time evolution of the quadrature operators \cite{Su15,Uk10}.
In particular, by manipulating the quadrature operators judiciously, we will provide an intuitive and efficient way
for establishing the measurement sequence and the corrective operations for the teleportation. This will simplify the
calculation significantly and make the analysis for teleportation involving large clusters manageable tasks.

As an orientation, let us consider a simple case with a linear four-mode cluster \cite{Me06}
as depicted in Fig.~\ref{fig:L4-1}. In view of the CZ-gate transformation \eref{cz_map}, the teleportation
that we wish to accomplish here amounts to implementing the mapping between the quadrature operators
of the output modes $\mu$, $\nu$ and the input modes $\alpha$, $\beta$
\begin{eqnarray}
\left(
  \begin{array}{c}
    \hat{q}_\mu\\
    \hat{p}_\mu\\
  \end{array}
\right)
\rightarrow
\left(
  \begin{array}{c}
    \hat{q}_\alpha \\
    \hat{p}_\alpha + \hat{q}_\beta\\
  \end{array}
\right)
\quad \mbox{and} \quad
\left(
  \begin{array}{c}
    \hat{q}_\nu\\
    \hat{p}_\nu\\
  \end{array}
\right)
\rightarrow
\left(
  \begin{array}{c}
    \hat{q}_\beta \\
    \hat{p}_\beta + \hat{q}_\alpha\\
  \end{array}
\right) \, .
\label{cz_goal}
\end{eqnarray}
Since the input modes $\alpha$, $\beta$ are coupled to the cluster nodes $1$, $4$, respectively,
by way of QND-gates, it follows from \eref{cz_map} that after these couplings, we have
\begin{eqnarray}
\left(
  \begin{array}{c}
     \hat{q}_\rho'\\
     \hat{p}_\rho'\\
  \end{array}
\right)
=
\left(
  \begin{array}{c}
    \hat{q}_\rho\\
    \hat{p}_\rho + \hat{q}_k\\
  \end{array}
\right)
\quad\mbox{and}\quad
\left(
  \begin{array}{c}
     \hat{q}_k'\\
     \hat{p}_k'\\
  \end{array}
\right)
=
\left(
  \begin{array}{c}
     \hat{q}_k\\
     \hat{p}_k + \hat{q}_\rho\\
  \end{array}
\right)\, ,
\label{in_coup}
\end{eqnarray}
where the subscripts $(\rho,k)=(\alpha,1)$ and $(\beta,4)$.
To accomplish the mapping \eref{cz_goal}, we note that the output mode $\mu$ is
obtained by correcting the final state of node 2 in the teleportation. Thus, in anticipation of
$\hop{q}{2}\rightarrow\hop{q}{\alpha}$ after corrective operations, we write the following
trivial identity making use of the $\hat{p}_1'$ entry of \eref{in_coup} (i.e., setting
$(\rho,k)=(\alpha,1)$ there and using the entry involving $\hop{p}{1}'$)
\begin{eqnarray}
\hat{q}_2 &=& \hat{q}_2 + (\hat{p}_1' - \hat{p}_1 - \hat{q}_\alpha)
\nonumber \\
&=& - \hat{q}_\alpha + \hat{p}_1'+(\hat{q}_2 - \hat{p}_1) \, ,
\label{q_L4-1}
\end{eqnarray}
where we have grouped the terms that would constitute a noise operator of \eref{noise_can} in reaching the second line.
Rearranging terms in the final identity above, one can write accordingly
\begin{eqnarray}
\hat{q}_\mu &\equiv& - \hat{q}_2 + \hat{p}_1'
\nonumber \\
&=& \hat{q}_\alpha + (\hat{p}_1 - \hat{q}_2) \, .
\label{qmu_L4-1}
\end{eqnarray}
Here we have written from \eref{q_L4-1} by collecting the image $\hop{q}{\alpha}$ of the intended mapping
$\hop{q}{2}\rightarrow\hop{q}{\alpha}$ and the noise operator $\hat{\delta}_1=(\hat{p}_1 - \hat{q}_2)$ to the
same side of the equation and relegating the rest to the other side, which is redefined as $\hop{q}{\mu}$.
Immediately, we see from this result that
by flipping the phase of the position operator $\hop{q}{2}$ for node 2 and then displacing
in accordance with the measurement outcome for $\hat{p}_1'$, one would be able to produce an output mode $\mu$
with $\hat{q}_\mu$ that differs from the input quadrature $\hat{q}_\alpha$ by just the excess noise
$\hat{\delta}_1$.

Similarly, anticipating $\hop{p}{2}\rightarrow\hop{p}{\alpha}+\hop{q}{\beta}$ subject to corrections,
one can make use of the $\hat{p}_\alpha'$ and the $\hat{p}_4'$ entries of \eref{in_coup}
and write down the following trivial identity for the momentum operator of node 2
\begin{eqnarray}
\hat{p}_2 &=& \hat{p}_2 + (\hat{p}_\alpha' - \hat{p}_\alpha - \hat{q}_1) + (\hat{p}_4' - \hat{p}_4 - \hat{q}_\beta )
\nonumber \\
&=& - (\hat{p}_\alpha + \hat{q}_\beta) + \hat{p}_\alpha' + \hat{p}_4' + (\hat{p}_2 - \hat{q}_1 - \hat{q}_3) - (\hat{p}_4 - \hat{q}_3) \, .
\label{p_L4-1}
\end{eqnarray}
Rearranging terms in the same way as in obtaining \eref{qmu_L4-1} from \eref{q_L4-1}, we get from the final identity in the last equation
\begin{eqnarray}
\hat{p}_\mu &\equiv& - \hat{p}_2 + \hat{p}_\alpha' + \hat{p}_4'
\nonumber \\
&=&  \hat{p}_\alpha + \hat{q}_\beta - (\hat{p}_2 - \hat{q}_1 - \hat{q}_3) + (\hat{p}_4 - \hat{q}_3) \, ,
\label{pmu_L4-1}
\end{eqnarray}
where the terms in the parentheses in the last expression are the noise operators $\hat{\delta}_2$ and $\hat{\delta}_4$ of \eref{noise_can}.
For the other output mode $\nu$, due to the symmetry between the modes in Fig.~\ref{fig:L4-1}, one can obtain
similar equations by exchanging the indices $1\leftrightarrow 4$, $2\leftrightarrow 3$, and $\alpha\leftrightarrow \beta$ in
\eref{qmu_L4-1} and \eref{pmu_L4-1}, and arrive at
\begin{eqnarray}
\hat{q}_\nu &\equiv& - \hat{q}_3 + \hat{p}_4'
\nonumber \\
&=& \hat{q}_\beta + (\hat{p}_4- \hat{q}_3) \, ,
\nonumber\\
\hat{p}_\nu &\equiv& - \hat{p}_3 + \hat{p}_\beta' + \hat{p}_1'
\nonumber \\
&=&  \hat{p}_\beta + \hat{q}_\alpha - (\hat{p}_3 - \hat{q}_2 - \hat{q}_4) + (\hat{p}_1 - \hat{q}_2) \, .
\label{nu_L4-1}
\end{eqnarray}

As the first line in each of the results \eref{qmu_L4-1}, \eref{pmu_L4-1}, and \eref{nu_L4-1} suggests, the CZ-gate
teleportation here requires measurements for the quadrature operators
$\hop{p}{\alpha}'$, $\hop{p}{\beta}'$, $\hop{p}{1}'$, and $\hop{p}{4}'$. Suppose the measurement results
are, respectively, $s_\alpha$, $s_\beta$, $s_1$, and $s_4$, the first lines of \eref{qmu_L4-1}, \eref{pmu_L4-1}, and \eref{nu_L4-1}
indicate that the corrective operations necessary for nodes 2 and 3 are
\begin{eqnarray}
\hop{X}{2}(s_1)\,\hop{Z}{2}(s_\alpha+s_4)\,\hop{F}{2}^2 \,\hop{X}{3}(s_4)\,\hop{Z}{3}(s_\beta+s_1)\,\hop{F}{3}^2 \, .
\label{corr_L4-1}
\end{eqnarray}
Here $\hop{X}{k}\equiv e^{-2i\hop{p}{k}s}$ and $\hop{Z}{k}\equiv e^{2i\hop{q}{k}s}$ are the Weyl-Heisenberg operators,
and $\hop{F}{k}\equiv e^{i\frac{\pi}{2}\hop{a}{k}^\dagger\hop{a}{k}}$ is the Fourier transform operator for mode $k$
that rotates its quadrature operators clockwise by an angle $(\pi/2)$ over the corresponding
(quantum mechanical) phase space \cite{Ko10,Fu11}.
It should be noted that due to the finitely squeezed cluster state, the teleported CZ-gate is imperfect. As one can
read off from the second lines
of \eref{qmu_L4-1}, \eref{pmu_L4-1}, and \eref{nu_L4-1}, here the quadrature operators for the output modes
differ from the ideal results (i.e., the image of the mapping in \eref{cz_goal}) by excess-noise contributions.

The example considered above is a relatively simple one, which, nevertheless, already shows the simplicity
of our Heisenberg approach.
To prepare for more elaborate tasks, let us explain in greater details
the tricks behind the manipulations of quadrature operators using the case of $\hop{q}{\mu}$ above.
Firstly, as pointed out earlier, one must notice that in the desired mapping \eref{cz_goal},
$\hop{q}{\mu}$ is constructed out of $\hop{q}{2}$. Thus, we start by writing
\begin{eqnarray}
\hop{q}{2}=\hop{q}{2}+\cdots \, ,
\label{q2_guess}
\end{eqnarray}
where the dots represent identities added from the entries of the
input-coupling equation \eref{in_coup} which serve for
two purposes: (i) they must include the image of $\hop{q}{\mu}$ for the
mapping \eref{cz_goal}, which is $\hop{q}{\alpha}$ here; (ii) they must
help ``eliminate" $\hop{q}{2}$ on the right-hand side of \eref{q2_guess} by
forming combination of terms that constitutes a nullifier of \eref{noise_can}.
Clearly, of the $\hop{q}{\alpha}'$ and $\hop{p}{1}'$ entries of \eref{in_coup} that
contain $\hop{q}{\alpha}$,
it is the latter that shall meet both demands (i) and (ii) here. It is then easy to arrive at \eref{q_L4-1},
and then \eref{qmu_L4-1} accordingly.
Similarly, the construction for $\hop{p}{\mu}$ in \eref{pmu_L4-1} can be achieved in the
same manner, except that two identities from the input-coupling equation \eref{in_coup} and two nullifiers
of \eref{noise_can} are invoked in this case.

\begin{figure}
\includegraphics*[width=75mm]{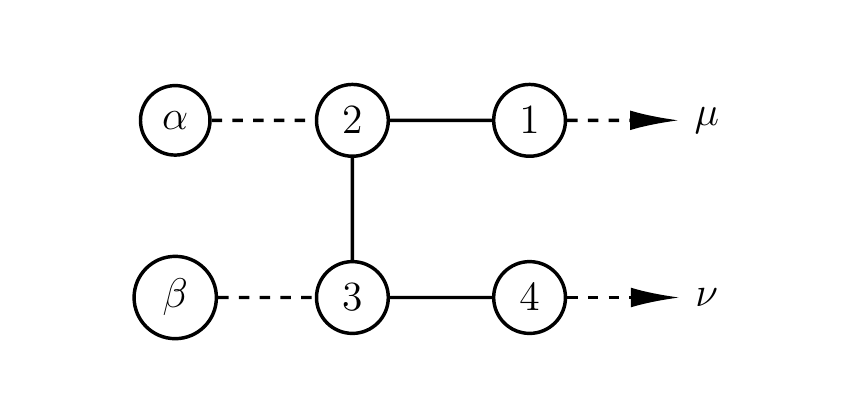}
\caption{CZ-gate teleportation using a linear four-mode cluster
with a node arrangement different from that of Fig.~\ref{fig:L4-1}.
\label{fig:L4-2}}
\end{figure}

As it turns out, when applying these tricks to different cluster structures and input-mode couplings,
there exist four major categories:
\begin{itemize}
\item[(a)] trivial cases: for which the quadrature manipulations can be carried out easily;
\item[(b)] FT-input cases: for which the quadrature manipulations require Fourier transformations
over the input modes prior to input couplings;
\item[(c)] FT-output cases: for which the quadrature manipulations require Fourier transformations
over the output cluster nodes at the output stage;
\item[(d)] FT-cluster cases: for which the quadrature manipulations require Fourier transformations over
some or all of the cluster nodes prior to input couplings.
\end{itemize}
It should be noted that these four categories are not mutually exclusive, nor is it unique for
the association of any gate-teleportation to these categories. For instance, we shall now illustrate with a
case which can be treated as a mixture of categories (b) and (c), or purely that of (d).
Let us consider again CZ-gate teleportation using a canonical linear four-mode cluster, but now with
different input-mode connections as depicted in Fig.~\ref{fig:L4-2}. As one can check through
the prescriptions above,
the quadrature manipulations in this case are no longer trivial \cite{note3}, and one must
consider possible scenarios with categories (b), (c), and (d) above.

After a few trial and errors, one can find that the CZ-gate teleportation in
the arrangement of Fig.~\ref{fig:L4-2} can be achieved by first applying inverse
Fourier transforms to the input modes $\alpha$, $\beta$ before they are coupled
to the cluster nodes:
\begin{eqnarray}
\left(
  \begin{array}{c}
    \hop{q}{\rho} \\
    \hop{p}{\rho} \\
  \end{array}
\right)
\xrightarrow{\hop{F}{\rho}^\dag}
\left(
  \begin{array}{c}
    \hop{q}{\rho}' \\
    \hop{p}{\rho}' \\
  \end{array}
\right)
=
\left(
  \begin{array}{r}
    \hop{p}{\rho} \\
    -\hop{q}{\rho} \\
  \end{array}
\right)
\label{F_in}
\end{eqnarray}
with $\rho=\alpha, \beta$. Subsequently, the Fourier-transformed input modes are
coupled to the cluster nodes 2, 3 via QND-gates and yield (cf.~\eref{in_coup})
\begin{eqnarray}
\left(
  \begin{array}{c}
     \hat{q}_\rho''\\
     \hat{p}_\rho''\\
  \end{array}
\right)
&=&
\left(
  \begin{array}{c}
    \hat{q}_\rho'\\
    \hat{p}_\rho' + \hat{q}_k\\
  \end{array}
\right)
=
\left(
  \begin{array}{c}
    \hat{p}_\rho\\
    -\hat{q}_\rho + \hat{q}_k\\
  \end{array}
\right) \, ,
\nonumber \\
\left(
  \begin{array}{c}
     \hat{q}_k'\\
     \hat{p}_k'\\
  \end{array}
\right)
&=&
\left(
  \begin{array}{c}
     \hat{q}_k\\
     \hat{p}_k + \hat{q}_\rho'\\
  \end{array}
\right)
=
\left(
  \begin{array}{c}
     \hat{q}_k\\
     \hat{p}_k + \hat{p}_\rho\\
  \end{array}
\right) \, ,
\label{in_coup2}
\end{eqnarray}
where the subscripts are $(\rho,k)=(\alpha,2)$ and $(\beta,3)$.
Like previously, in view of the CZ-gate mapping \eref{cz_goal} and that the output mode
$\mu$ corresponds to the corrected cluster mode 1, one can proceed similarly to what was done in \eref{q_L4-1}.
However, things are a little more intricate here. It transpires that
Fourier transformations over the output cluster nodes are also necessary when producing the output modes here.
Therefore, for node 1, in place of $\hop{q}{1}\rightarrow\hop{q}{\alpha}$, we have here
$\hop{p}{1}\rightarrow\hop{q}{\alpha}$ when supplemented with corrections.
According to the prescriptions (i) and (ii) described after \eref{q2_guess}, we can make use of
the $\hop{p}{\alpha}''$ entry of \eref{in_coup2} and write
\begin{eqnarray}
\hat{p}_1 &=& \hat{p}_1 + (\hat{p}_\alpha'' + \hat{q}_\alpha - \hop{q}{2})
\nonumber \\
&=& \hat{q}_\alpha + \hat{p}_\alpha''+(\hat{p}_1 - \hat{q}_2) \, .
\label{q_L4-2}
\end{eqnarray}
The last equality thus suggests
\begin{eqnarray}
\hat{q}_\mu &\equiv& \hat{p}_1 - \hat{p}_\alpha''
\nonumber \\
&=& \hat{q}_\alpha + (\hat{p}_1 - \hat{q}_2) \,
\label{qmu_L4-2}
\end{eqnarray}
with the terms in the parentheses the noise operator $\hop{\delta}{1}$.
Similarly, due to the Fourier transformation, instead of $\hop{p}{1}\rightarrow (\hop{p}{\alpha}+\hop{q}{\beta})$,
we expect here
$-\hop{q}{1}\rightarrow (\hop{p}{\alpha}+\hop{q}{\beta})$ when corrections are applied.
We can thus write by way of the $\hat{p}_2'$ and the $\hat{p}_\beta''$ entries of \eref{in_coup2}
\begin{eqnarray}
\hat{q}_1 &=& \hat{q}_1 + (\hat{p}_2' - \hop{p}{2} - \hat{p}_\alpha) - (\hat{p}_\beta'' + \hat{q}_\beta - \hat{q}_3 )
\nonumber \\
&=& - (\hat{p}_\alpha + \hat{q}_\beta) - \hat{p}_\beta'' + \hat{p}_2' - (\hat{p}_2 - \hat{q}_1 - \hat{q}_3) \, .
\label{p_L4-2}
\end{eqnarray}
Rearranging terms in the last identity above thus yields
\begin{eqnarray}
\hat{p}_\mu &\equiv& - \hat{q}_1 - \hat{p}_\beta'' + \hat{p}_2'
\nonumber \\
&=&  \hat{p}_\alpha + \hat{q}_\beta + (\hat{p}_2 - \hat{q}_1 - \hat{q}_3) \, ,
\label{pmu_L4-2}
\end{eqnarray}
where the terms in the parentheses in the last expression is the noise operator $\hat{\delta}_2$.

As previously, the symmetry in the arrangement of Fig.~\ref{fig:L4-2} allows us to write
down the formulas for the output mode $\nu$ easily from \eref{qmu_L4-2} and \eref{pmu_L4-2}.
We find
\begin{eqnarray}
\hat{q}_\nu &\equiv& \hat{p}_4 - \hat{p}_\beta''
\nonumber \\
&=& \hat{q}_\beta + (\hat{p}_4- \hat{q}_3) \, ,
\nonumber\\
\hat{p}_\nu &\equiv& - \hat{q}_4 - \hat{p}_\alpha'' + \hat{p}_3'
\nonumber \\
&=&  \hat{p}_\beta + \hat{q}_\alpha + (\hat{p}_3 - \hat{q}_2 - \hat{q}_4) \, .
\label{nu_L4-2}
\end{eqnarray}
Therefore, according to the first lines of the results in \eref{qmu_L4-2}, \eref{pmu_L4-2},
and \eref{nu_L4-2}, the CZ-gate teleportation in the setting of Fig.~\ref{fig:L4-2} requires measurements for the quadrature operators
$\hop{p}{\alpha}''$, $\hop{p}{\beta}''$, $\hop{p}{2}'$, and $\hop{p}{3}'$. If the corresponding
measurement outcomes are, respectively, $s_\alpha$, $s_\beta$, $s_2$, and $s_3$, the first lines
of \eref{qmu_L4-2}, \eref{pmu_L4-2}, and \eref{nu_L4-2} indicate that the corrective operations
needed for the cluster modes 1 and 4 are
\begin{eqnarray}
\hop{X}{1}^\dag(s_\alpha)\,\hop{Z}{1}^\dag(s_\beta-s_2)\,\hop{F}{1}^\dag\,
\hop{X}{4}^\dag(s_\beta)\,\hop{Z}{4}^\dag(s_\alpha-s_3)\,\hop{F}{4}^\dag \, .
\label{corr_L4-2}
\end{eqnarray}
This furnishes the scenario for the CZ-gate teleportation using a canonical linear four-mode cluster
in the arrangement of Fig.~\ref{fig:L4-2}. We see that the quadrature manipulations here
have invoked the Fourier transformations stipulated for categories (b) and (c).
As pointed out earlier, one can also treat this case following the scheme of category (d);
this is illustrated in Appendix \ref{sec:FTc}.

Although the Heisenberg approach presented above does seem less systematic than the Schr\"{o}dinger approach \cite{Me06}.
However, with the limited number of categories listed above, usually a few trial and errors can quickly
bring forward the correct teleportation scenarios, and the calculation requires much less algebra than the Schr\"{o}dinger
approach does. This is especially appealing when large clusters are
involved in the teleportation process, such as the cases that we will study in the following section.

We have so far been focusing on establishing the procedures for
CZ-gate teleportation through given cluster designs. An important
issue next is then how one can quantify the quality of the
teleported CZ-gate. Since CZ-gate is capable of entangling its two
input modes, we will quantify the quality of the teleported gate
through the entanglement in the output modes when non-entangled
input modes are supplied. Although entanglement quantification for
general CV states remains a major challenge \cite{BrvL05,Ad07}, the
teleported states that will concern us in this work belong entirely
to the class of two-mode Gaussian states. Their entanglement
properties can be fully quantified through partial transposition of
their covariance matrices \cite{Si00,WW01}. This is based on the
fact that for any Gaussian state all symplectic eigenvalues of its
covariance matrix cannot be less than $1/4$
\cite{BrvL05,note35}. For a two-mode Gaussian state, suppose the
partial transposition of its covariance matrix has symplectic
eigenvalues $\lambda_\pm$ with $\lambda_-\le\lambda_+$. Since
$\lambda_+ \ge 1/4$ always \cite{Se04}, here the physicality of the
partially transposed covariance matrix is determined entirely by the
smaller symplectic eigenvalue $\lambda_-$, and a measure for the
degree of entanglement for the state is provided by the logarithmic
negativity (or log-negativity, for short) \cite{VW02}
\begin{eqnarray}
E_N = - \ln(\min\{1,4\lambda_-\}) \, ,
\label{EN}
\end{eqnarray}
where the function $\min(x,y)$ yields the smaller of $x$ and $y$. For entangled two-mode Gaussian states, violation
of physicality upon partial transposition is then signaled by $\lambda_- < 1/4$, which results
in non-zero, positive $E_N$ according to \eref{EN}.

If we denote the output quadratures of the teleported CZ-gate in the form of a column vector
$\hat{\xi} \equiv (\hat{q}_\mu,\hat{p}_\mu,\hop{q}{\nu},\hop{p}{\nu})^T$.
The covariance matrix for the output modes $\mu$, $\nu$ is then a real,
symmetric $4\times 4$ matrix with elements \cite{WM08}
\begin{eqnarray}
V_{kl} \equiv \frac{1}{2}\,\left\langle \left\{\Delta\hat{\xi}_k,\Delta\hat{\xi}_l\right\} \right\rangle\, ,
\label{Vkl}
\end{eqnarray}
where $k,l=1\sim 4$, $\Delta\hop{\xi}{k}\equiv \hop{\xi}{k} - \langle\hop{\xi}{k}\rangle$, and $\{A,B\}\equiv AB+BA$.
Since we are using the Heisenberg picture for the time evolution, the expectation values
here are evaluated with respect to the initial state of the system.
As it emerges, the covariance matrices for the output modes $\mu$, $\nu$ for all cases considered in this
work share an ``X-form"
\begin{eqnarray}
V = \left(
                \begin{array}{cccc}
                      a & 0 & 0 & c \\
                      0 & b & c & 0     \\
                      0 & c & a & 0     \\
                      c & 0 & 0 & b
                \end{array}
          \right)
\label{V_X}
\end{eqnarray}
with, according to \eref{Vkl},
\begin{eqnarray}
a = \left\langle (\Delta\hop{q}{\sigma})^2 \right\rangle \, , \quad
b = \left\langle (\Delta\hop{p}{\sigma})^2 \right\rangle \, , \quad
c = \left\langle \{\Delta\hop{q}{\sigma},\Delta\hop{p}{\sigma'}\} \right\rangle/2 \, ,
\label{abc}
\end{eqnarray}
where $\sigma, \sigma'=\mu,\nu$, and $\sigma'\neq\sigma$ in the expression for $c$.
The partial transposition for the covariance matrix \eref{V_X}
with respect to, say, the output mode $\nu$ is effected through replacing
in \eref{abc} $\hat{q}_\nu\rightarrow +\hat{q}_\nu$ and
$\hat{p}_\nu\rightarrow -\hat{p}_\nu$ (that is, ``time reversal" in mode $\nu$ \cite{Si00}).
The symplectic eigenvalues of the partially transposed covariance matrix are then found to be \cite{Se04}
\begin{eqnarray}
\lambda_\pm = |\sqrt{ab} \pm c| \, .
\label{symp}
\end{eqnarray}
Substituting the expression for $\lambda_-$ into \eref{EN}, we can thus write the log-negativity
\begin{eqnarray}
E_N = \max\left\{0,- \ln(4\,|\sqrt{ab}-c|)\right\}
\label{EN2}
\end{eqnarray}
with $\max\{x,y\}$ being the larger of $x$ and $y$.

As an illustration, let us look back at the CZ-gate teleported
in the arrangement of Fig.~\ref{fig:L4-2} with a canonical linear four-mode cluster. Suppose
the input modes $\alpha$, $\beta$ are independent coherent states, one then has
$\left\langle (\Delta\hat{q}_\rho)^2\right\rangle = \left\langle (\Delta\hat{p}_\rho)^2\right\rangle = 1/4$
and $\left\langle \{\Delta\hop{q}{\rho},\Delta\hop{p}{\rho}\} \right\rangle = 0$
for both $\rho=\alpha,\beta$. If the cluster state had been prepared from momentum-squeezed vacuum states with
identical squeezing parameter $r$, so that $\langle \hop{\delta}{k}^2 \rangle = e^{-2r}/4$ for all $k=1\sim 4$
in \eref{noise_can}. Using the results \eref{qmu_L4-2}, \eref{pmu_L4-2}, \eref{nu_L4-2}, one can obtain
the quadrature correlators \eref{abc}
\begin{eqnarray}
a = \frac{(1+e^{-2r})}{4} \, , \quad
b = \frac{(2+e^{-2r})}{4} \, , \quad
c = \frac{1}{4} \, .
\label{abc_L4}
\end{eqnarray}
It then follows from \eref{EN2} that the log-negativity here reads
\begin{eqnarray}
E_{N,L4} = \max\left\{0,- \ln\!\left(\sqrt{(1+e^{-2r})(2+e^{-2r})}-1\right)\right\}\, ,
\label{EN_L4}
\end{eqnarray}
where the additional subscript ``$L4$" indicates the ``linear
four-mode". One can check that for
$r<\ln(\frac{\sqrt{\sqrt{17}+3}}{2})\simeq 0.29$ the log-negativity
$E_{N,L4}$ vanishes identically. In other words, the presence of the
excess noise in the teleported CZ-gate here has entirely corrupted
the entanglement in the output modes for finite range of squeezing
levels. We will examine in the following section how the multi-rail
design shown in Fig.~\ref{fig:MR_tel} can help reduce the excess
noise, and hence improve entanglement in the output modes of the
teleported CZ-gate.

\section{\label{sec:CZ} Controlled-phase (CZ) gate teleportation using multi-rail clusters}
\begin{figure}
\includegraphics*[width=81mm]{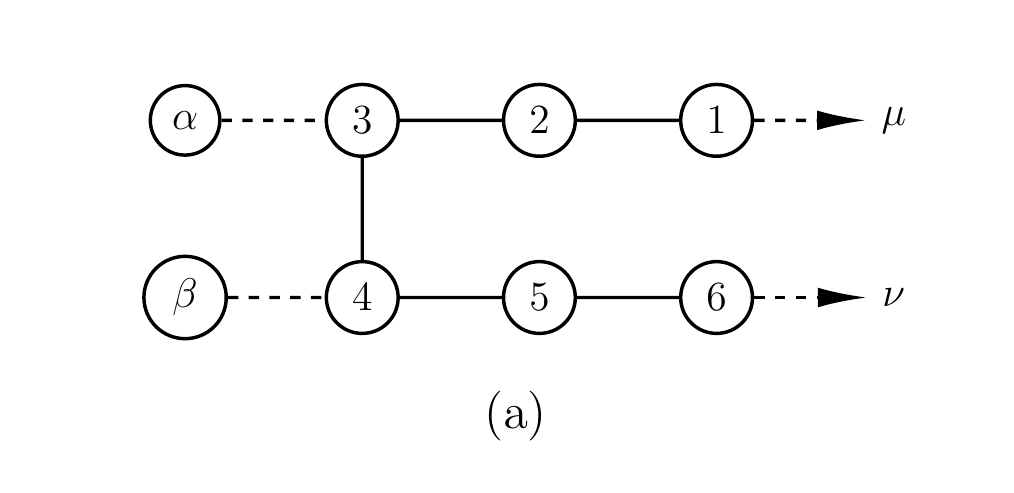}
\includegraphics*[width=81mm]{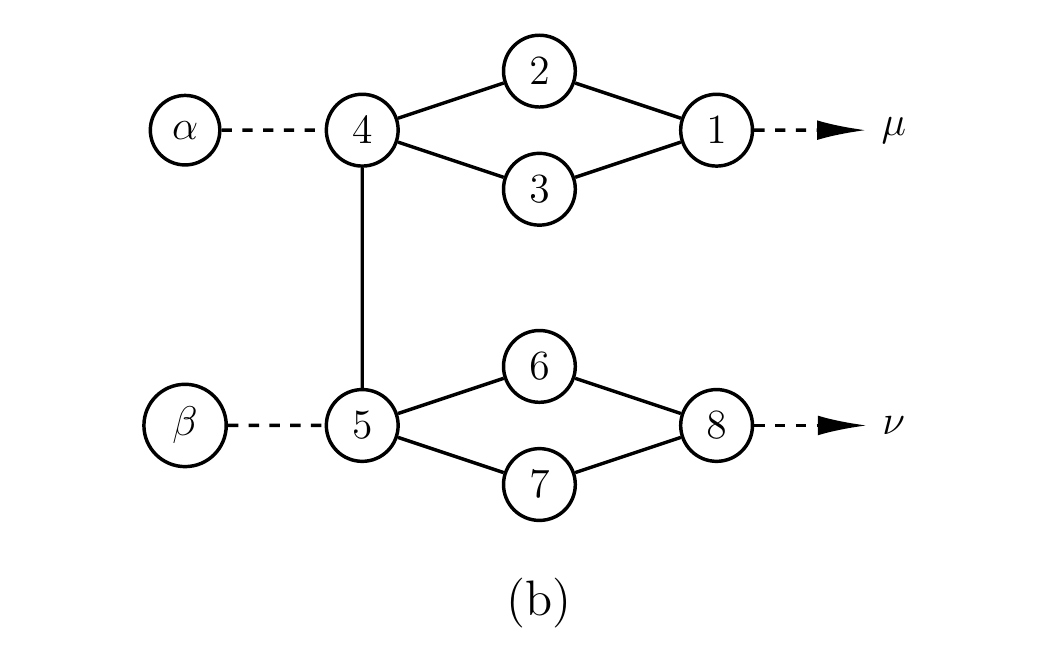}
\includegraphics*[width=81mm]{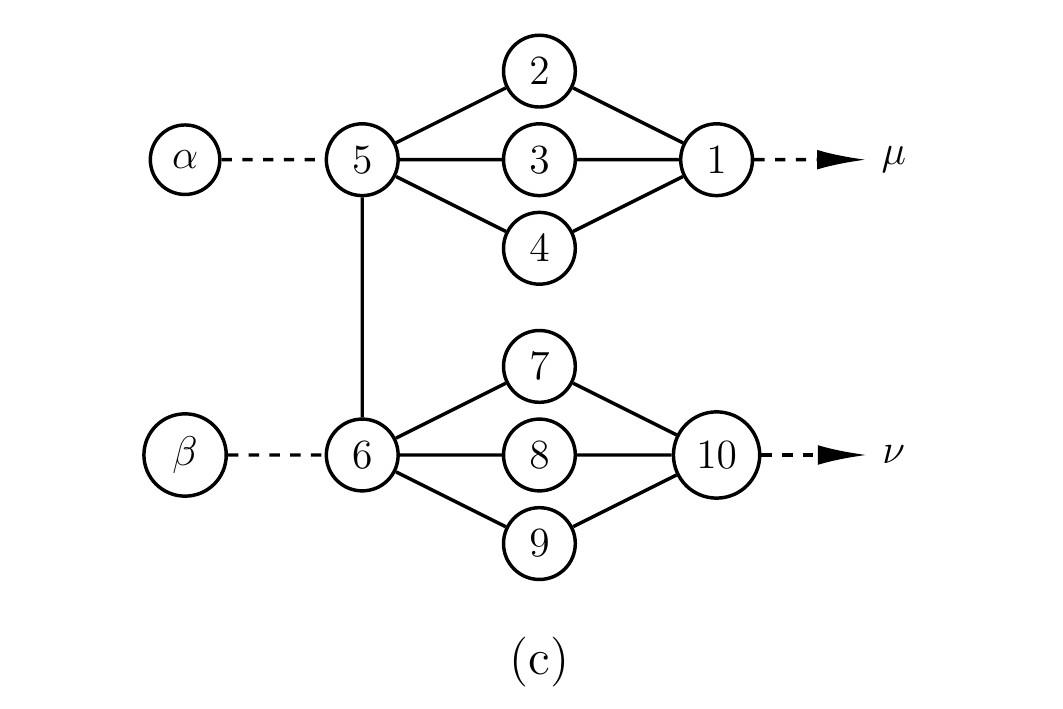}
\includegraphics*[width=81mm]{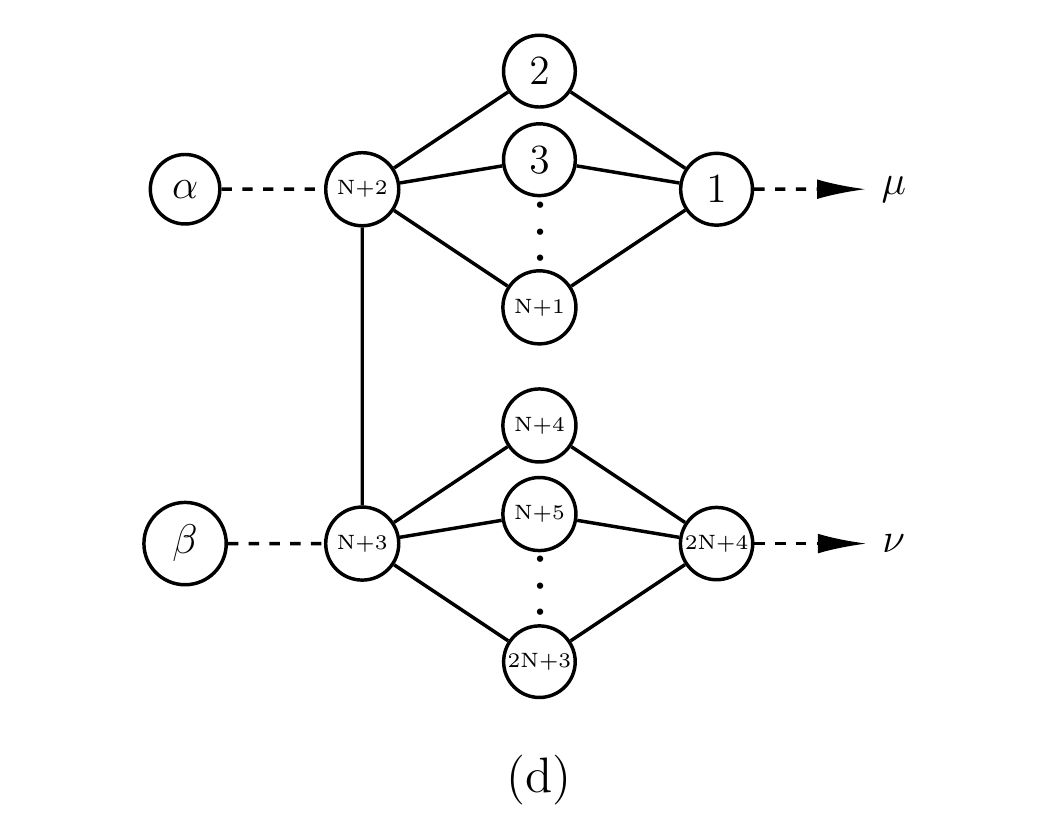}
\caption{CZ-gate teleportation using clusters with multi-rail designs of (a) single-rail,
(b) two-rail, (c) three-rail, and (d) $N$-rail structures. Namely, the ``teleporting arms"
(i.e., the segments of nodes 3--2--1 and 4--5--6) of the single-rail cluster in (a) are
replaced with multi-rail structures in (b)--(d) according to the designs of
Fig.~\ref{fig:MR_tel}. Here the dashed lines and the arrows have the same meanings as
those in Figs.~\ref{fig:L4-1} and \ref{fig:L4-2}.
\label{fig:MR_cz}}
\end{figure}

The noise-reduction scheme through multi-rail designs shown in Fig.~\ref{fig:MR_tel} was originally
proposed in Ref.~\onlinecite{vL07} for the teleportation of single-qumode gates.
For the CZ-gate, since there are two input qumodes, it is necessary to implement the
multi-rail structure for each of the ``teleporting arms", such as the segments of nodes 2--1
and nodes 3--4 in the case of Fig.~\ref{fig:L4-2}. As is clear from Fig.~\ref{fig:MR_tel}, the
minimal cluster for the multi-rail design requires three nodes for each arm. Therefore, for CZ-gate
teleportation we will start with a linear six-mode cluster, and then increase the complexity of the
cluster by building the multi-rail structure into the arms as shown in Fig.~\ref{fig:MR_cz}.

In what follows, we will consider two classes of resource states for the CZ-gate teleportation.
The first is the canonical cluster states that are fabricated through QND-gates, as we have already
considered in the previous section. The second will be a class of cluster states proposed by
van Loock and coworkers \cite{vL07} which can be constructed using linear-optical networks and thus
will be termed the linear-optical cluster states in the following. As we will see, these two classes
of cluster states possess different noise structures, and hence are not identical at finite squeezing.
However, in the ideal, infinite-squeezing limit, they become completely identical
since all excess noise tend to zero in this limit. Therefore, at finite squeezing, we expect to
see different entanglement properties in the teleported CZ-gates when different classes of
cluster states are employed for the teleportation, as we shall now investigate.

\subsection{\label{sec:canon} Canonical cluster states}
Let us start by considering CZ-gate teleportation using a canonical linear six-mode cluster as illustrated
in Fig.~\ref{fig:MR_cz}(a), which corresponds to a ``single-rail" design.
The quadrature manipulations in this case are very similar to those for the linear four-mode cluster
of Fig.~\ref{fig:L4-2} demonstrated in the previous section. Again, it is necessary to subject the input modes $\alpha$, $\beta$
to the Fourier transformations \eref{F_in} before they are coupled with the cluster nodes.
Subsequent QND-couplings between the Fourier-transformed input modes $\alpha'$, $\beta'$ and
the cluster nodes 3, 4, respectively, thus yield again \eref{in_coup2}
with here the sets of subscripts $(\rho,k)=(\alpha,3)$ and $(\beta,4)$.
Making use of these coupling equations and the nullifiers for the
linear six-mode cluster, one can obtain in the same manner as before
\begin{eqnarray}
\hat{q}_\mu &\equiv& -\hat{q}_1 - \hat{p}_\alpha'' + \hop{p}{2}
\nonumber \\
&=& \hat{q}_\alpha + (\hat{p}_2 - \hat{q}_1 - \hop{q}{3}) \, ,
\nonumber \\
\hat{p}_\mu &\equiv& - \hat{p}_1 - \hat{p}_\beta'' + \hat{p}_3'
\nonumber \\
&=&  \hat{p}_\alpha + \hat{q}_\beta - (\hop{p}{1}-\hop{q}{2})
+ (\hat{p}_3 - \hat{q}_2 - \hat{q}_4) \, ,
\label{mu_L6}
\end{eqnarray}
and
\begin{eqnarray}
\hat{q}_\nu &\equiv& -\hat{q}_6 - \hat{p}_\beta'' + \hop{p}{5}
\nonumber \\
&=& \hat{q}_\beta + (\hat{p}_5- \hat{q}_4 - \hop{q}{6}) \, ,
\nonumber\\
\hat{p}_\nu &\equiv& - \hat{p}_6 - \hat{p}_\alpha'' + \hat{p}_4'
\nonumber \\
&=&  \hat{p}_\beta + \hat{q}_\alpha - (\hop{p}{6}-\hop{q}{5})
+ (\hat{p}_4 - \hat{q}_3 - \hat{q}_5) \, .
\label{nu_L6}
\end{eqnarray}
Accordingly, from the first line in each of the results for the output quadratures above,
we see that here the CZ-gate teleportation calls for measurements over the momentum operators
$\hop{p}{\alpha}''$, $\hop{p}{\beta}''$, $\hop{p}{2}$, $\hop{p}{3}'$, $\hop{p}{4}'$, and $\hop{p}{5}$.
Suppose the corresponding measurement outcomes are, respectively, $s_\alpha$, $s_\beta$, $s_2$, $s_3$,
$s_4$, and $s_5$, the gate teleportation can then be achieved upon correcting the cluster nodes 1 and 6
with the operations
\begin{eqnarray}
\hop{X}{1}(s_2-s_\alpha)\,\hop{Z}{1}(s_3-s_\beta)\,\hop{F}{1}^2\,\hop{X}{6}(s_5-s_\beta)\,\hop{Z}{6}(s_4-s_\alpha)\,\hop{F}{6}^2 \, ,
\label{corr_L6}
\end{eqnarray}
which have been read off from \eref{mu_L6} and \eref{nu_L6} in the same way as
detailed previously for the cases of linear four-mode clusters.

We next turn to the multi-rail variants of the linear six-mode cluster.
As explained in the beginning of this section, in order to implement the multi-rail design
of Fig.~\ref{fig:MR_tel} into the linear six-mode cluster in Fig.~\ref{fig:MR_cz}(a), we
replace each of the teleporting arms (i.e., the segments 3--2--1 and 4--5--6)
of the cluster with multi-rail structures as in Figs.~\ref{fig:MR_cz}(b)--(d).
For the two-rail cluster of Fig.~\ref{fig:MR_cz}(b), we find that the CZ-gate teleportation
can again be achieved by first applying the Fourier transformations \eref{F_in} over the input modes
$\alpha$, $\beta$ before they are coupled to the cluster. After that, QND-couplings between the
Fourier transformed input modes $\alpha'$, $\beta'$ and the cluster nodes 4, 5, respectively,
again lead to \eref{in_coup2} with the subscripts now $(\rho,k)=(\alpha,4)$ and $(\beta,5)$.
With the help of these coupling equations and the nullifiers for the
present cluster, it is straightforward to establish the following through the same tricks as before
\begin{eqnarray}
\hat{q}_\mu &\equiv& -\hat{q}_1 - \hat{p}_\alpha'' + \frac{\hop{p}{2}+\hop{p}{3}}{2}
\nonumber \\
&=& \hat{q}_\alpha
+ \frac{1}{2} \left[(\hat{p}_2 - \hat{q}_1 - \hop{q}{4}) + (\hat{p}_3 - \hat{q}_1 - \hop{q}{4}) \right]\, ,
\nonumber \\
\hat{p}_\mu &\equiv& - \hat{p}_1 - \hat{p}_\beta'' + \hat{p}_4'
\nonumber \\
&=&  \hat{p}_\alpha + \hat{q}_\beta - (\hop{p}{1}-\hop{q}{2}-\hop{q}{3})
+ (\hat{p}_4 - \hat{q}_2 - \hat{q}_3 - \hop{q}{5}) \, ,
\label{mu_2R}
\end{eqnarray}
and
\begin{eqnarray}
\hat{q}_\nu &\equiv& -\hat{q}_8 - \hat{p}_\beta'' + \frac{\hop{p}{6}+\hop{p}{7}}{2}
\nonumber \\
&=& \hat{q}_\beta
+ \frac{1}{2} \left[(\hat{p}_6 - \hat{q}_5 - \hop{q}{8}) + (\hat{p}_7 - \hat{q}_5 - \hop{q}{8}) \right]\, ,
\nonumber\\
\hat{p}_\nu &\equiv& - \hat{p}_8 - \hat{p}_\alpha'' + \hat{p}_5'
\nonumber \\
&=&  \hat{p}_\beta + \hat{q}_\alpha - (\hop{p}{8}-\hop{q}{6}-\hop{q}{7})
+ (\hat{p}_5 - \hat{q}_4 - \hat{q}_6 - \hop{q}{7}) \, .
\label{nu_2R}
\end{eqnarray}
Here, however, it should be noted that in \eref{mu_2R} and \eref{nu_2R}, one can in general set for
$\hop{q}{\mu}$ and $\hop{q}{\nu}$
\begin{eqnarray}
\hat{q}_\mu &\equiv& -\hat{q}_1 - \hat{p}_\alpha'' + (\eta_1\,\hop{p}{2}+\eta_2\,\hop{p}{3})
\nonumber \\
&=& \hat{q}_\alpha
+ \left[\eta_1\,(\hat{p}_2 - \hat{q}_1 - \hop{q}{4}) +\eta_2\,(\hat{p}_3 - \hat{q}_1 - \hop{q}{4}) \right] \, ,
\nonumber \\
\hat{q}_\nu &\equiv& -\hat{q}_8 - \hat{p}_\beta'' + (\eta_3\,\hop{p}{6}+\eta_4\,\hop{p}{7})
\nonumber \\
&=& \hat{q}_\beta
+ \left[\eta_3\,(\hat{p}_6 - \hat{q}_5 - \hop{q}{8}) +\eta_4\,(\hat{p}_7 - \hat{q}_5 - \hop{q}{8}) \right] \, ,
\label{qq_2R}
\end{eqnarray}
where $\eta_1+\eta_2=1$ and $\eta_3+\eta_4=1$, so that in each equation the first line would always be identical to the second.
Clearly, since both $(\hop{q}{\mu}-\hop{q}{\alpha})$ and
$(\hop{q}{\nu}-\hop{q}{\beta})$ in \eref{qq_2R} remain to yield linear combinations of excess-noise operators, these
choices of $\hop{q}{\mu}$ and $\hop{q}{\nu}$ should also serve well for the CZ-gate teleportation.
However, as one can prove easily, for equally squeezed initial cluster modes, the symmetrical arrangement
$\eta_k=1/2$ for all $k=1\sim 4$ in \eref{qq_2R}
(i.e., corresponding to \eref{mu_2R} and \eref{nu_2R}) would minimize the excess noise in the teleported
$\hop{q}{\mu}$ and $\hop{q}{\nu}$. This is also the case for the three-rail design that we shall discuss shortly, and
is in fact the reason behind the noise reduction in this multi-rail approach. For instance, the choice $\eta_{1,3}=1$
and $\eta_{2,4}=0$ in \eref{qq_2R} would yield $\hop{q}{\mu}$ and $\hop{q}{\nu}$ identical to those for the
single-rail cluster in \eref{mu_L6} and \eref{nu_L6}, respectively, except for change of indices in the nodes.
In this case, there would thus be no any noise reduction in the teleportation despite the implemented multi-rail structure.

For the case of the three-rail cluster illustrated in Fig.~\ref{fig:MR_cz}(c), the calculation proceeds almost identically
to that for the two-rail case above, apart from modifications in the node indices. Instead of repeating
similar details, here we list only the results that we find
\begin{eqnarray}
\hat{q}_\mu &\equiv& -\hat{q}_1 - \hat{p}_\alpha'' + \frac{\hop{p}{2}+\hop{p}{3}+\hop{p}{4}}{3}
\nonumber \\
&=& \hat{q}_\alpha
+ \frac{1}{3} \left[(\hat{p}_2 - \hat{q}_1 - \hop{q}{5}) + (\hat{p}_3 - \hat{q}_1 - \hop{q}{5})
+ (\hat{p}_4 - \hat{q}_1 - \hop{q}{5}) \right]\, ,
\nonumber \\
\hat{p}_\mu &\equiv& - \hat{p}_1 - \hat{p}_\beta'' + \hat{p}_5'
\nonumber \\
&=&  \hat{p}_\alpha + \hat{q}_\beta - (\hop{p}{1}-\hop{q}{2}-\hop{q}{3}-\hop{q}{4})
+ (\hat{p}_5 - \hat{q}_2 - \hat{q}_3 - \hop{q}{4} - \hop{q}{6}) \, ,
\label{mu_3R}
\end{eqnarray}
and
\begin{eqnarray}
\hat{q}_\nu &\equiv& -\hat{q}_{10} - \hat{p}_\beta'' + \frac{\hop{p}{7}+\hop{p}{8}+\hop{p}{9}}{3}
\nonumber \\
&=& \hat{q}_\beta
+ \frac{1}{3} \left[(\hat{p}_7 - \hat{q}_6 - \hop{q}{10}) + (\hat{p}_8 - \hat{q}_6 - \hop{q}{10})
+ (\hat{p}_9 - \hat{q}_6 - \hop{q}{10}) \right]\, ,
\nonumber\\
\hat{p}_\nu &\equiv& - \hat{p}_{10} - \hat{p}_\alpha'' + \hat{p}_6'
\nonumber \\
&=&  \hat{p}_\beta + \hat{q}_\alpha - (\hop{p}{10}-\hop{q}{7}-\hop{q}{8}-\hop{q}{9})
+ (\hat{p}_6 - \hat{q}_5 - \hat{q}_7 - \hop{q}{8} - \hop{q}{9})  \, .
\label{nu_3R}
\end{eqnarray}
As for the two-rail case above, here one could in general put $\hop{q}{\mu}$ and $\hop{q}{\nu}$ in the form
\begin{eqnarray}
\hat{q}_\mu &\equiv& -\hat{q}_1 - \hat{p}_\alpha'' + (\eta_1\,\hop{p}{2}+\eta_2\,\hop{p}{3}+\eta_3\,\hop{p}{4})
\nonumber \\
&=& \hat{q}_\alpha
+ \left[\eta_1\,(\hat{p}_2 - \hat{q}_1 - \hop{q}{5}) +\eta_2\,(\hat{p}_3 - \hat{q}_1 - \hop{q}{5})
+\eta_3\,(\hat{p}_4 - \hat{q}_1 - \hop{q}{5}) \right] \, ,
\nonumber \\
\hat{q}_\nu &\equiv& -\hat{q}_{10} - \hat{p}_\beta'' + (\eta_4\,\hop{p}{7}+\eta_5\,\hop{p}{8}+\eta_6\,\hop{p}{9})
\nonumber \\
&=& \hat{q}_\beta
+ \left[\eta_4\,(\hat{p}_7 - \hat{q}_6 - \hop{q}{10}) +\eta_5\,(\hat{p}_8 - \hat{q}_6 - \hop{q}{10})
+\eta_6\,(\hat{p}_9 - \hat{q}_6 - \hop{q}{10}) \right] \, ,
\label{qq_3R}
\end{eqnarray}
where $\eta_1+\eta_2+\eta_3=1$ and $\eta_4+\eta_5+\eta_6=1$, so that each identity would always hold.
Nevertheless, one can again show easily that it is
the symmetrical choice $\eta_k=1/3$ for all $k=1\sim 6$ of \eref{mu_3R} and \eref{nu_3R}
that would minimize the excess noise in the teleported $\hop{q}{\mu}$ and $\hop{q}{\nu}$ had the cluster
been prepared from uniformly squeezed vacuum states.
We will prove below that this result can indeed be generalized to arbitrary $N$-rail design for
the canonical cluster.

With the foregoing results, it is straightforward to extend the consideration to a design with an arbitrary
number of rail. For the $N$-rail canonical cluster in Fig.~\ref{fig:MR_cz}(d),
from Eqs.~\eref{mu_L6}, \eref{nu_L6}, \eref{mu_2R}, \eref{nu_2R}, \eref{mu_3R}, and \eref{nu_3R},
one can write down inductively for the output modes of the teleported CZ-gate
\begin{eqnarray}
\hat{q}_\mu &=& \hat{q}_\alpha + \frac{1}{N} \sum_{k=2}^{N+1}\hop{\delta}{k} \, ,
\nonumber\\
\hat{p}_\mu &=& \hat{p}_\alpha + \hop{q}{\beta} - \hop{\delta}{1} + \hop{\delta}{N+2} \, ,
\nonumber\\
\hat{q}_\nu &=& \hat{q}_\beta + \frac{1}{N} \sum_{k=N+4}^{2N+3}\hop{\delta}{k} \, ,
\nonumber\\
\hat{p}_\nu &=& \hat{p}_\beta + \hop{q}{\alpha} - \hop{\delta}{2N+4} + \hop{\delta}{N+3} \, .
\label{qp_NR}
\end{eqnarray}
Note that here we have omitted the parts of the equations that would reveal the measurement sequence and corrective
operations for the teleportation (namely, corresponding to the first lines of the results \eref{mu_L6},
\eref{nu_L6}, and etc.) This is because here we are concerned primarily with the entanglement properties of the
output modes, and thus need only the parts of the equations listed in \eref{qp_NR}. Also note that we have
expressed the formulas here in terms of the noise operators $\hop{\delta}{k}$ instead of the quadrature operators,
as this will be more convenient for evaluating the quadrature correlators \eref{abc}.

Like previously for the case of Fig.~\ref{fig:L4-2} with a canonical linear four-mode cluster,
we shall suppose that the input modes $\alpha$, $\beta$ here are independent coherent states and that
the cluster nodes are uniformly squeezed with squeezing parameter $r$. Moreover, according
to \eref{noise_can}, the noise operators $\hop{\delta}{k}$ for canonical cluster states are independent from each
other. We can thus find from \eref{qp_NR} the quadrature correlators \eref{abc} for the output modes of
the teleported CZ-gate
\begin{eqnarray}
a = \frac{(1+\frac{1}{N}e^{-2r})}{4} \, , \quad
b = \frac{(1+e^{-2r})}{2} \, , \quad
c = \frac{1}{4} \, .
\label{abc_NR}
\end{eqnarray}
It should be noted that the correlators $b$ and $c$ are both $N$-independent here.
One can now obtain immediately from \eref{EN2} the log-negativity for the output modes
\begin{eqnarray}
E_{N,NR} = \max\left\{0,- \ln\!\left(\sqrt{2(1+\frac{e^{-2r}}{N})(1+e^{-2r})}-1\right)\right\} \,
\label{EN_NR}
\end{eqnarray}
with the subscript ``$NR$" standing for ``$N$-rail". Note that here
the result for the two-rail ($N=2$) case is identical to
that for the linear four-mode cluster \eref{EN_L4} in the previous section. We will defer
further analysis for the result \eref{EN_NR} until the end of the
next subsection, where comparisons between the results from two
classes of resource states will be made. Before closing this
subsection, let us look back at the expressions for $\hop{q}{\mu}$
and $\hop{q}{\nu}$ in \eref{qp_NR}. Like previously for the two-rail
and the three-rail cases, the coefficient $1/N$ in front of the
summation over the noise operators $\hop{\delta}{k}$ in fact
corresponds to the optimal choice for reducing the excess noise in
the teleported $\hop{q}{\mu}$ and $\hop{q}{\nu}$ when the cluster is
uniformly squeezed. A generic expression for $\hop{q}{\mu}$ and $\hop{q}{\nu}$
in the present $N$-rail teleportation takes the form
\begin{eqnarray}
\hat{q}_\sigma = \hat{q}_\rho + \sum_k \eta_k\,\hop{\delta}{k} \, ,
\label{q_gen_NR}
\end{eqnarray}
where $(\sigma,\rho)=(\mu,\alpha)$ or $(\nu,\beta)$, and the summation has the same range as
in \eref{qp_NR}. For the same reason as in \eref{qq_2R} and \eref{qq_3R}, here we must
constraint the coefficients $\eta_k$ such that $\sum_k \eta_k =1$.
Since the noise operators $\hop{\delta}{k}$ are independent from each other, the excess noise
in the teleported $\hop{q}{\sigma}$ is simply
$\langle(\hop{q}{\sigma}-\hop{q}{\rho})^2\rangle=\sum_k \eta_k^2\,\langle\hop{\delta}{k}^2\rangle$. For
uniformly squeezed clusters $\langle\hop{\delta}{k}^2\rangle$ has the same value for all $k$, thus
the minimization for the excess noise is the same as that for the sum $\sum_k\eta_k^2$. Geometrically,
this amounts to finding the point on the $N$-dimensional hyperplane $\sum_k \eta_k=1$ which has the shortest (Euclidean)
distance to the origin. The answer is clearly the ``symmetric point" with $\eta_k=1/N$ for all $k$.
A rigorous derivation for this result is elementary, for instance, based on the method of Lagrange multipliers.
We note that for $\hop{q}{\sigma}$ in \eref{q_gen_NR} the summation over $k$ covers all mid-rail nodes
in each set of the multi-rails, i.e., nodes $k=2\sim (N+1)$ for the set with $\sigma=\mu$ and $k=(N+4)\sim(2N+3)$
for the set with $\sigma=\nu$ in Fig.~\ref{fig:MR_cz}(d). Thus, the symmetric arrangement $\eta_k=1/N$ in \eref{q_gen_NR}
corresponds to distributing the excess noise to each of the multi-rail with equal weight.

\subsection{\label{sec:lo} Linear-optical cluster states}
The construction of canonical cluster states requires implementation of QND-gates, which can be
experimentally challenging especially when the cluster is large and the qumodes are encoded through
spatially separated optical modes. In this context, van Loock and coworkers proposed in
Ref.~\onlinecite{vL07} an alternative route to preparing optical cluster states through
offline squeezers and linear-optical networks of beam-splitters and phase shifters.
In essence, one has a set of spatially separated optical modes that are squeezed locally and
sent through an optical network of passive elements, which is capable of entangling these modes
and setting up correlations among their quadrature operators that are akin to those in canonical clusters.
In particular, despite the more complicated noise operators $\hop{\delta}{k}$ in this case (see below),
in the limit of infinite squeezing, just like canonical cluster states, all $\hop{\delta}{k}$ would
tend to zero and the state would approach an ideal cluster state. These states have thus been dubbed
``linear-optical cluster states" in the present paper.

In a linear-optical cluster state, the cluster correlations among
its nodes are established through a unitary transformation associated with the action of a
linear-optical network \cite{Br05,Re94} over a set of offline squeezed initial modes.
If the annihilation operators for these initial modes are
$\hop{\bar{a}}{l}$, the linear-optical network induces the transformation
\begin{eqnarray}
\hop{a}{k} = \sum_l U_{kl}\,\hop{\bar{a}}{l} \, ,
\label{U_transf}
\end{eqnarray}
where $U_{kl}$ is the $k,l$ element of the corresponding unitary matrix $U$, and $\hop{a}{k}$ is the
annihilation operator for the resultant cluster mode $k$.
In pursuance of furnishing the cluster correlations, the $k$-th row $\vec{u}_k$ of $U$ must take the particular form \cite{vL07}
\begin{eqnarray}
\vec{u}_k = \vec{\alpha}_k + i \sum_{l\in N_k} \vec{\alpha}_l \, ,
\label{r_vec}
\end{eqnarray}
where $\vec{\alpha}_k$ are real row vectors and $N_k$ indicates the set of nodes in the cluster that are connected to
node $k$. Unitarity of the matrix $U$ entails orthonormality of the row vectors $\vec{u}_k$, which yields
\begin{eqnarray}
&\vec{\alpha}_k\,\vec{\alpha}_l^{\,T} + \sum_{m\in N_k, n\in N_l} \vec{\alpha}_m\,\vec{\alpha}_n^{\,T} = \delta_{kl} \, ,
\nonumber \\
&\vec{\alpha}_k\,\sum_{n\in N_l} \vec{\alpha}_n^{\,T} -  \sum_{m\in N_k} \vec{\alpha}_m\,\vec{\alpha}_l^{\,T} = 0 \, ,
\label{geo_con}
\end{eqnarray}
where $T$ signifies matrix transposition and $\delta_{kl}$ is Kronecker's delta.
For any given geometry of the cluster, one can obtain the conditions $\vec{\alpha}_k$ (or, in fact, their inner products) must satisfy
in accordance with \eref{geo_con}.
Writing
\begin{eqnarray}
G_{kl} \equiv \vec{\alpha}_k\,\vec{\alpha}_l^{\,T} \,,
\label{Gkl}
\end{eqnarray}
one can solve easily (though often tediously) from \eref{geo_con} the values of $G_{kl}$ for all $k,l$.
Since $G_{kl}$ is symmetric in $k$ and $l$, a cluster with $M$ nodes would have $M(M+1)/2$
such ``geometric constraints" for the $\vec{\alpha}_k$'s.
Following these constraints, one can construct $\vec{\alpha}_k$ accordingly, and hence the row vectors
$\vec{u}_k$ of $U$ through \eref{r_vec}. Clearly, the choices for
$\vec{\alpha}_k$ and thus $U$ are not unique. As long as
\eref{geo_con} is satisfied, the matrix $U$ would always lead to the
desired cluster correlations for the linear-optical cluster state
\cite{vL07}. Details of these calculations for the
linear-optical cluster states that will concern us are summarized in
Appendix~\ref{sec:GU}. Here we just quote the expression for the
noise operators \cite{vL07}
\begin{eqnarray}
\hop{\delta}{k}
= \sum_l \left(\alpha_{k,l} + \sum_{m\in N_k} \sum_{n\in N_m} \alpha_{n,l} \right) \hop{\bar{p}}{l} \, ,
\label{noise_lo}
\end{eqnarray}
where $\hop{q}{k}$, $\hop{p}{k}$ are quadrature operators for node $k$, $\alpha_{k,l}$ denotes the
$l$-th component of the vector $\vec{\alpha}_k$, and,
as in \eref{noise_can}, $\hop{\bar{p}}{l}=e^{-r_l}\hop{p}{l}^{(0)}$ is the momentum operator for
the offline squeezed initial mode $l$. It is clear that here the noise operators have much more complicated
dependence on the initial mode operators than that in \eref{noise_can} for canonical cluster states.
Nevertheless, in the ideal limit of infinite squeezing (i.e., $r_l\rightarrow\infty$ for every $l$),
one would still recover the limit of ideal cluster states as previously for canonical cluster states.
Therefore, we see that the unitary transformation \eref{U_transf} with
row vectors of the form \eref{r_vec} does implement successfully the intended cluster correlations
among the nodes of the cluster.

Let us turn now to the study of CZ-gate teleportation through linear-optical cluster states.
As previously, we shall first illustrate the calculation utilizing the simple case of Fig.~\ref{fig:L4-2} with
a linear four-mode cluster of the linear-optical type. In order to reduce the demand for squeezing resources \cite{Br05}, here the coupling
between the input modes $\alpha$, $\beta$ and the cluster nodes will be achieved through beam-splitter couplings
(via ``Bell measurements" \cite{Uk10}), instead of the QND-couplings previously. Specifically, for
the input mode $\rho=\alpha,\beta$ and the cluster node $k=2,3$ of Fig.~\ref{fig:L4-2}, the
coupling is effected through a 50:50 beam splitter, so that the annihilation operators for the modes
transform as \cite{Ko10,Fu11}
\begin{eqnarray}
\left(
  \begin{array}{c}
     \hat{a}_{\rho_1}\\
     \hat{a}_{\rho_2}\\
  \end{array}
\right)
=
\left(
  \begin{array}{cc}
    \frac{1}{\sqrt{2}} & \frac{1}{\sqrt{2}} \\
    \frac{1}{\sqrt{2}} & \frac{-1}{\sqrt{2}} \\
  \end{array}
\right)
\left(
  \begin{array}{c}
    \hat{a}_\rho\\
    \hat{a}_k\\
  \end{array}
\right) \, ,
\label{lo_in_coup}
\end{eqnarray}
where the subscripts are $(\rho,k)=(\alpha,2)$ and $(\beta,3)$, and hence
$(\rho_1,\rho_2) = (\alpha_1,\alpha_2)$ and $(\beta_1,\beta_2)$, correspondingly.
Namely, the beam splitter has coupled the input mode $\rho$ and the cluster mode $k$
in producing the modes $\rho_1$ and $\rho_2$, which have the quadrature operators,
according to \eref{lo_in_coup},
\begin{eqnarray}
\left(
  \begin{array}{c}
     \hat{q}_{\rho_1}\\
     \hat{p}_{\rho_1}\\
  \end{array}
\right)
=
\left(
  \begin{array}{c}
     \frac{\hop{q}{\rho}+\hop{q}{k}}{\sqrt{2}}\\
     \frac{\hop{p}{\rho}+\hop{p}{k}}{\sqrt{2}}\\
  \end{array}
\right)
\quad \mbox{and} \quad
\left(
  \begin{array}{c}
     \hat{q}_{\rho_2}\\
     \hat{p}_{\rho_2}\\
  \end{array}
\right)
=
\left(
  \begin{array}{c}
     \frac{\hop{q}{\rho}-\hop{q}{k}}{\sqrt{2}}\\
     \frac{\hop{p}{\rho}-\hop{p}{k}}{\sqrt{2}}\\
  \end{array}
\right) \, .
\label{r1_r2}
\end{eqnarray}

For CZ-gate teleportation, like previously for canonical clusters, again we wish to achieve the mapping \eref{cz_goal}
through quadrature-manipulation tricks by utilizing
the input-coupling equation \eref{r1_r2} and the cluster correlations \eref{noise_lo}.
However, it should be noted that previously the input coupling \eref{in_coup} through QND-gate had led to
trivial transformations in both $\hop{q}{\rho}'$ and $\hop{q}{k}'$ entries (as they belong to the
``controlled" part of the QND operation \eref{cz_map}). It was thus quite obvious that one must employ the
$\hop{p}{\rho}'$ and $\hop{p}{k}'$ entries there for quadrature manipulations, and consequently the measurement sequence
involves invariably the $\hat{p}'$s and never the $\hat{q}'$s \cite{note4}.
Here, the situation is more complicated since all entries of \eref{r1_r2}
are non-trivial and extra care is needed in determining the entries to be used (and hence
the quadratures to be measured for the teleportation).

For instance, in the case of Fig.~\ref{fig:L4-2} with a linear-optical cluster, anticipating that
$\hop{q}{1}\rightarrow\hop{q}{\alpha}$ after
corrective operations, one would write according to points (i) and (ii) listed below \eref{q2_guess}
\begin{eqnarray}
\hop{q}{1} &=& \hop{q}{1} + \hop{q}{\alpha} + \cdots
\nonumber\\
&=& \hop{q}{\alpha} + \hop{q}{1} - \hop{p}{2} + \hop{q}{3} + \cdots \, ,
\label{q1_lo}
\end{eqnarray}
where the dots in the second line remain to be fixed through entries
of the input-coupling equation \eref{r1_r2}. However, it is clear
from \eref{r1_r2} that the added $\hop{p}{2}$ must appear together
with $\hop{p}{\alpha}$, no matter whether it is $\hop{p}{\alpha_1}$
or $\hop{p}{\alpha_2}$ that will be measured. Similarly, according
to \eref{r1_r2}, the added term $\hop{q}{3}$ in \eref{q1_lo} must
bring along $\hop{q}{\beta}$ to the equation. It is thus evident that
this scheme will not work and one must attempt with different
scenarios. For instance, if one attempt instead with
$\hop{p}{1}\rightarrow \hop{q}{\alpha}$ up to corrective operations,
one can write using the $\hop{q}{\alpha_2}$ entry of \eref{r1_r2}
\begin{eqnarray}
\hat{p}_1 &=& \hat{p}_1 - (\sqrt{2}\hop{q}{\alpha_2}-\hop{q}{\alpha}+\hop{q}{2})
\nonumber \\
&=& \hat{q}_\alpha - \sqrt{2} \hat{q}_{\alpha_2}+(\hat{p}_1 - \hat{q}_2) \, .
\label{q_L4_lo}
\end{eqnarray}
Recognizing the nullifier $(\hat{p}_1 - \hat{q}_2)=\hop{\delta}{1}$ in the last
line, one can thus assign
\begin{eqnarray}
\hat{q}_\mu &\equiv& \hat{p}_1 + \sqrt{2}\hat{q}_{\alpha_2}
\nonumber \\
&=& \hat{q}_\alpha + (\hat{p}_1 - \hat{q}_2) \, .
\label{qmu_L4_lo}
\end{eqnarray}
This result indicates that Fourier transformation is needed here and thus suggests
for the output quadrature $\hop{p}{\mu}$ the mapping
$\hop{q}{1}\rightarrow - (\hop{p}{\alpha}+\hop{q}{\beta})$ subject to corrections.
With this observation, we write utilizing the $\hat{p}_{\alpha_1}$ and
the $\hat{q}_{\beta_2}$ entries of \eref{r1_r2}
\begin{eqnarray}
\hat{q}_1 &=& \hat{q}_1 + (\sqrt{2}\,\hat{p}_{\alpha_1} - \hop{p}{\alpha} - \hat{p}_2) + (\sqrt{2}\,\hat{q}_{\beta_2} - \hat{q}_\beta + \hat{q}_3 )
\nonumber \\
&=& -(\hat{p}_\alpha + \hat{q}_\beta) + \sqrt{2} (\hat{p}_{\alpha_1} + \hat{q}_{\beta_2}) - (\hat{p}_2 - \hat{q}_1 - \hat{q}_3) \, .
\label{p_L4_lo}
\end{eqnarray}
Rearranging terms in the last expression yields immediately the desired equation
\begin{eqnarray}
\hat{p}_\mu &\equiv& - \hat{q}_1 + \sqrt{2} (\hat{p}_{\alpha_1} + \hat{q}_{\beta_2})
\nonumber \\
&=&  \hat{p}_\alpha + \hat{q}_\beta + (\hat{p}_2 - \hat{q}_1 - \hat{q}_3)
\label{pmu_L4_lo}
\end{eqnarray}
with the terms in the parentheses in the last line the nullifier $\hat{\delta}_2$.
Exploiting the symmetry among the modes in Fig.~\ref{fig:L4-2}, based on the
results \eref{qmu_L4_lo} and \eref{pmu_L4_lo}, one can write immediately for
the other output mode $\nu$
\begin{eqnarray}
\hat{q}_\nu &\equiv& \hat{p}_4 + \sqrt{2}\hat{q}_{\beta_2}
\nonumber \\
&=& \hat{q}_\beta + (\hat{p}_4 - \hat{q}_3) \, ,
\nonumber\\
\hat{p}_\nu &\equiv& - \hat{q}_4 + \sqrt{2} (\hat{p}_{\beta_1} + \hat{q}_{\alpha_2})
\nonumber \\
&=&  \hat{p}_\beta + \hat{q}_\alpha + (\hat{p}_3 - \hat{q}_2 - \hat{q}_4) \, .
\label{nu_L4_lo}
\end{eqnarray}
From the second line in each of the equations in \eref{qmu_L4_lo}, \eref{pmu_L4_lo}, and \eref{nu_L4_lo},
we see that we have
achieved the intended mapping \eref{cz_goal} for the CZ-gate, except for the presence of excess-noise
terms $\hop{\delta}{k}$ due to finite squeezing. One can also read off from the first lines of these equations
that the CZ-gate teleportation here requires measurements over the quadrature operators
$\hop{p}{\alpha_1}$, $\hop{q}{\alpha_2}$, $\hop{p}{\beta_1}$, and $\hop{q}{\beta_2}$. For measurement
outcomes with, respectively, $s_{\alpha_1}$, $s_{\alpha_2}$, $s_{\beta_1}$, and $s_{\beta_2}$, one can find
from the first lines of \eref{qmu_L4_lo}, \eref{pmu_L4_lo}, and \eref{nu_L4_lo} that the corrective
operations are here
\begin{eqnarray}
\hop{X}{1}(\sqrt{2}\,s_{\alpha_2})\,\hop{Z}{1}(\sqrt{2}\,(s_{\alpha_1}+s_{\beta_2}))\,\hop{F}{1}^\dag\,
\hop{X}{4}(\sqrt{2}\,s_{\beta_2})\,\hop{Z}{4}(\sqrt{2}\,(s_{\beta_1}+s_{\alpha_2}))\,\hop{F}{4}^\dag \, .
\label{corr_L4_lo}
\end{eqnarray}

To quantify the quality of the teleported CZ-gate, we calculate as ever the log-negativity \eref{EN2}
for the output modes $\mu$ and $\nu$. For this purpose, it is necessary
to find first the explicit forms for the noise operators $\hop{\delta}{k}$, so that one can calculate
the quadrature correlators \eref{abc} using \eref{qmu_L4_lo}, \eref{pmu_L4_lo}, and \eref{nu_L4_lo}.
Using the matrix $U$ found for linear four-mode clusters in Appendix \ref{sec:GU}, we get
\begin{eqnarray}
\hop{\delta}{1} &=& \sqrt{2} \,\hop{\bar{p}}{1} \, ,
\nonumber\\
\hop{\delta}{2} &=& \frac{5}{\sqrt{10}} \,\hop{\bar{p}}{2} + \frac{1}{\sqrt{2}} \,\hop{\bar{p}}{4} \, ,
\nonumber\\
\hop{\delta}{3} &=& \frac{1}{\sqrt{2}} \,\hop{\bar{p}}{1} + \frac{5}{\sqrt{10}} \,\hop{\bar{p}}{3} \, ,
\nonumber\\
\hop{\delta}{4} &=& \sqrt{2} \,\hop{\bar{p}}{4} \,.
\label{delta_L4}
\end{eqnarray}
As noted following \eref{noise_lo}, the noise operators here have
much more complicated forms than their canonical-cluster counterparts \eref{noise_can}. This will also
be seen for other linear-optical cluster states that we will consider later.

For the sake of calculating the quadrature correlators \eref{abc} explicitly, as before,
we shall consider identically squeezed cluster modes
with squeezing parameter $r$ and independent coherent-state input modes. For the noise
operators, we note that previously for canonical cluster states only the
auto-correlators $\langle\hop{\delta}{k}^2\rangle$ are non-vanishing,
while here, according to \eref{delta_L4}, there can be non-zero
cross correlators $\langle\hop{\delta}{k}\,\hop{\delta}{l}\rangle$ with $k\neq l$.
With this precaution, it is then not difficult to find the quadrature correlators \eref{abc} by means of
\eref{qmu_L4_lo}, \eref{pmu_L4_lo}, \eref{nu_L4_lo}, together with \eref{delta_L4} and get
\begin{eqnarray}
a = \frac{(1+2e^{-2r})}{4} \, , \quad
b = \frac{(2+3e^{-2r})}{4} \, , \quad
c = \frac{(1+e^{-2r})}{4} \, .
\label{abc_L4_lo}
\end{eqnarray}
Substituting these results into \eref{EN2}, we find the log-negativity for the output modes
\begin{eqnarray}
E_{N,L4'} = \max\left\{0,-\ln\!\left(\sqrt{(1+2e^{-2r})(2+3e^{-2r})}-(1+e^{-2r})\right)\right\} \, ,
\label{EN_L4_lo}
\end{eqnarray}
where the prime over the subscript $L4$ has been added to tell the cluster from its canonical counterpart earlier
(cf.~\eref{EN_L4}).
As with canonical clusters, here excess noise due to finite squeezing has
impaired the entanglement between the output modes $\mu$ and $\nu$. In particular,
one can check easily that the entanglement \eref{EN_L4_lo} in the output modes attained
here is worse than that of \eref{EN_L4} obtained previously
for canonical linear four-mode cluster. It is therefore of interest to
examine also the performance of the multi-rail noise
reduction for teleportation with linear-optical clusters.
Before delving into this analysis, on the grounds of the foregoing discussions,
we note that the calculations for CZ-gate teleportation using linear-optical clusters
are indeed very similar to those for canonical cluster states,
except for the differences in the input coupling and the correlations in the noise operators.
We will therefore be very brief with the calculations for each individual case in the following
and take them merely as intermediate steps leading to the general results for CZ-gate
teleportation through an arbitrary $N$-rail linear-optical cluster.

Let us start with the linear six-mode cluster of Fig.~\ref{fig:MR_cz}(a), where the input modes are
now coupled to the cluster through two 50:50 beam splitters as in \eref{lo_in_coup}.
The input coupling between the input modes $\alpha$, $\beta$ and the cluster nodes 3, 4, respectively,
yields the consequential modes given by \eref{r1_r2} with $(\rho,k)=(\alpha,3)$ and $(\beta,4)$.
Making use of these coupling equations, the same tricks as those illustrated earlier for the linear four-mode
cluster produce the results
\begin{eqnarray}
\hat{q}_\mu &\equiv& \hop{q}{1} - \hat{p}_2 + \sqrt{2}\hat{q}_{\alpha_1}
\nonumber \\
&=& \hat{q}_\alpha - (\hat{p}_2 - \hat{q}_1 - \hop{q}{3}) \, ,
\nonumber\\
\hat{p}_\mu &\equiv& \hat{p}_1 + \sqrt{2} (\hat{p}_{\alpha_2} + \hat{q}_{\beta_1})
\nonumber \\
&=&  \hat{p}_\alpha + \hat{q}_\beta + (\hop{p}{1} - \hop{q}{2}) - (\hat{p}_3 - \hat{q}_2 - \hat{q}_4) \, ,
\label{mu_L6_lo}
\end{eqnarray}
and
\begin{eqnarray}
\hat{q}_\nu &\equiv& \hop{q}{6} - \hat{p}_5 + \sqrt{2}\hat{q}_{\beta_1}
\nonumber \\
&=& \hat{q}_\beta - (\hat{p}_5 - \hat{q}_4 - \hop{q}{6}) \, ,
\nonumber\\
\hat{p}_\nu &\equiv& \hat{p}_6 + \sqrt{2} (\hat{q}_{\alpha_1} + \hat{p}_{\beta_2})
\nonumber \\
&=&  \hat{p}_\beta + \hat{q}_\alpha + (\hop{p}{6} - \hop{q}{5}) - (\hat{p}_4 - \hat{q}_3 - \hat{q}_5) \, .
\label{nu_L6_lo}
\end{eqnarray}
By means of the matrix $U$ found for linear six-mode clusters in Appendix \ref{sec:GU}, one can arrive at
the noise operators from \eref{noise_lo}
\begin{eqnarray}
\hop{\delta}{1} &=& \sqrt{2} \,\hop{\bar{p}}{1} \, ,
\nonumber\\
\hop{\delta}{2} &=& \sqrt{3} \,\hop{\bar{p}}{2} \, ,
\nonumber\\
\hop{\delta}{3} &=& \frac{1}{\sqrt{2}} \,\hop{\bar{p}}{1} + \sqrt{\frac{13}{6}} \,\hop{\bar{p}}{3}
+ \frac{1}{\sqrt{3}} \,\hop{\bar{p}}{5} \, ,
\nonumber\\
\hop{\delta}{4} &=& \frac{1}{\sqrt{3}} \,\hop{\bar{p}}{2} + \sqrt{\frac{13}{6}} \,\hop{\bar{p}}{4}
+ \frac{1}{\sqrt{2}} \,\hop{\bar{p}}{6} \, ,
\nonumber\\
\hop{\delta}{5} &=& \sqrt{3} \,\hop{\bar{p}}{5} \, ,
\nonumber\\
\hop{\delta}{6} &=& \sqrt{2} \,\hop{\bar{p}}{6} \, .
\label{delta_L6}
\end{eqnarray}

For the two-rail cluster of Fig.~\ref{fig:MR_cz}(b), again beam-splitter coupling \eref{lo_in_coup}
between the input modes $\alpha$, $\beta$ and the cluster nodes 4, 5, respectively,
generates the consequential modes given by \eref{r1_r2} with $(\rho,k)=(\alpha,4)$ and $(\beta,5)$.
These results and the quadrature-manipulation tricks lead to the output quadratures for the
teleported CZ-gate
\begin{eqnarray}
\hat{q}_\mu &\equiv& \hop{q}{1} - \frac{\hat{p}_2+\hop{p}{3}}{2} + \sqrt{2}\hat{q}_{\alpha_1}
\nonumber \\
&=& \hat{q}_\alpha
- \frac{1}{2} \left[(\hat{p}_2 - \hat{q}_1 - \hop{q}{4})+(\hat{p}_3 - \hat{q}_1 - \hop{q}{4})\right] \, ,
\nonumber\\
\hat{p}_\mu &\equiv& \hat{p}_1 + \sqrt{2} (\hat{p}_{\alpha_2} + \hat{q}_{\beta_1})
\nonumber \\
&=&  \hat{p}_\alpha + \hat{q}_\beta + (\hop{p}{1} - \hop{q}{2} -\hop{q}{3})
- (\hat{p}_4 - \hat{q}_2 - \hat{q}_3 -\hop{q}{5}) \, ,
\label{mu_2R_lo}
\end{eqnarray}
and
\begin{eqnarray}
\hat{q}_\nu &\equiv& \hop{q}{8} - \frac{\hat{p}_6+\hop{p}{7}}{2} + \sqrt{2}\hat{q}_{\beta_1}
\nonumber \\
&=& \hat{q}_\beta
- \frac{1}{2} \left[(\hat{p}_6 - \hat{q}_5 - \hop{q}{8})+(\hat{p}_7 - \hat{q}_5 - \hop{q}{8})\right] \, ,
\nonumber\\
\hat{p}_\nu &\equiv& \hat{p}_8 + \sqrt{2} (\hat{q}_{\alpha_1} + \hat{p}_{\beta_2})
\nonumber \\
&=&  \hat{p}_\beta + \hat{q}_\alpha + (\hop{p}{8} - \hop{q}{6} -\hop{q}{7})
- (\hat{p}_5 - \hat{q}_4 - \hat{q}_6 -\hop{q}{7}) \, .
\label{nu_2R_lo}
\end{eqnarray}
The noise operators \eref{noise_lo} for the two-rail linear-optical cluster can again be obtained
via the corresponding unitary matrix $U$ listed in in Appendix \ref{sec:GU}. We find
\begin{eqnarray}
\hop{\delta}{1} &=& \sqrt{3} \,\hop{\bar{p}}{1} \, ,
\nonumber\\
\hop{\delta}{2} &=& \sqrt{3} \,\hop{\bar{p}}{2} \, ,
\nonumber\\
\hop{\delta}{3} &=& \frac{2}{\sqrt{3}} \,\hop{\bar{p}}{2} + \sqrt{\frac{5}{3}} \,\hop{\bar{p}}{3} \, ,
\nonumber\\
\hop{\delta}{4} &=& \frac{2}{\sqrt{3}} \,\hop{\bar{p}}{1} + \sqrt{\frac{34}{15}} \,\hop{\bar{p}}{4} + \frac{1}{\sqrt{15}} \,\hop{\bar{p}}{6}
+ \frac{1}{\sqrt{3}} \,\hop{\bar{p}}{7} \, ,
\nonumber\\
\hop{\delta}{5} &=& \frac{1}{\sqrt{3}} \,\hop{\bar{p}}{2} + \frac{1}{\sqrt{15}} \,\hop{\bar{p}}{3} + \sqrt{\frac{34}{15}} \,\hop{\bar{p}}{5}
+ \frac{2}{\sqrt{3}} \,\hop{\bar{p}}{8}\, ,
\nonumber\\
\hop{\delta}{6} &=& \sqrt{\frac{5}{3}} \,\hop{\bar{p}}{6} + \frac{2}{\sqrt{3}} \,\hop{\bar{p}}{7} \, ,
\nonumber\\
\hop{\delta}{7} &=& \sqrt{3} \,\hop{\bar{p}}{7} \, ,
\nonumber\\
\hop{\delta}{8} &=& \sqrt{3} \,\hop{\bar{p}}{8} \, .
\label{delta_2R}
\end{eqnarray}

Finally, for the three-rail cluster of Fig.~\ref{fig:MR_cz}(c), the beam-splitter
coupling \eref{lo_in_coup} between the input modes $\alpha$, $\beta$ and, respectively, the cluster nodes 5, 6
yields the post-coupling modes of \eref{r1_r2} with $(\rho,k)=(\alpha,5)$ and $(\beta,6)$.
Immediately, the same tricks as previously bring forth
\begin{eqnarray}
\hat{q}_\mu &\equiv& \hop{q}{1} - \frac{\hat{p}_2+\hop{p}{3}+\hop{p}{4}}{3} + \sqrt{2}\hat{q}_{\alpha_1}
\nonumber \\
&=& \hat{q}_\alpha
- \frac{1}{3} \left[(\hat{p}_2 - \hat{q}_1 - \hop{q}{5})+(\hat{p}_3 - \hat{q}_1 - \hop{q}{5})
+(\hat{p}_4 - \hat{q}_1 - \hop{q}{5})\right] \, ,
\nonumber\\
\hat{p}_\mu &\equiv& \hat{p}_1 + \sqrt{2} (\hat{p}_{\alpha_2} + \hat{q}_{\beta_1})
\nonumber \\
&=&  \hat{p}_\alpha + \hat{q}_\beta + (\hop{p}{1} - \hop{q}{2} -\hop{q}{3}-\hop{q}{4})
- (\hat{p}_5 - \hat{q}_2 - \hat{q}_3 -\hop{q}{4}-\hop{q}{6}) \, ,
\label{mu_3R_lo}
\end{eqnarray}
and
\begin{eqnarray}
\hat{q}_\nu &\equiv& \hop{q}{10} - \frac{\hat{p}_7+\hop{p}{8}+\hop{p}{9}}{3} + \sqrt{2}\hat{q}_{\beta_1}
\nonumber \\
&=& \hat{q}_\beta
- \frac{1}{3} \left[(\hat{p}_7 - \hat{q}_6 - \hop{q}{10})+(\hat{p}_8 - \hat{q}_6 - \hop{q}{10})
+(\hat{p}_9 - \hat{q}_6 - \hop{q}{10})\right] \, ,
\nonumber\\
\hat{p}_\nu &\equiv& \hat{p}_{10} + \sqrt{2} (\hat{q}_{\alpha_1} + \hat{p}_{\beta_2})
\nonumber \\
&=&  \hat{p}_\beta + \hat{q}_\alpha + (\hop{p}{10} - \hop{q}{7} -\hop{q}{8}-\hop{q}{9})
- (\hat{p}_6 - \hat{q}_5 - \hat{q}_7 -\hop{q}{8}-\hop{q}{9}) \, .
\label{nu_3R_lo}
\end{eqnarray}
The $U$-matrix obtained in Appendix \ref{sec:GU} for the three-rail cluster enables us to find
the noise operators \eref{noise_lo} here explicitly
\begin{eqnarray}
\hop{\delta}{1} &=& 2 \,\hop{\bar{p}}{1} \, ,
\nonumber\\
\hop{\delta}{2} &=& \sqrt{3} \,\hop{\bar{p}}{2} \, ,
\nonumber\\
\hop{\delta}{3} &=& \frac{2}{\sqrt{3}} \,\hop{\bar{p}}{2} + \sqrt{\frac{5}{3}} \,\hop{\bar{p}}{3} \, ,
\nonumber\\
\hop{\delta}{4} &=& \frac{2}{\sqrt{3}} \,\hop{\bar{p}}{2} + \frac{2}{\sqrt{15}} \,\hop{\bar{p}}{3} + \sqrt{\frac{7}{5}} \,\hop{\bar{p}}{4} \, ,
\nonumber\\
\hop{\delta}{5} &=& \frac{3}{2} \,\hop{\bar{p}}{1} + \frac{1}{2}\sqrt{\frac{65}{7}} \,\hop{\bar{p}}{5} + \frac{1}{\sqrt{35}} \,\hop{\bar{p}}{7}
+ \frac{1}{\sqrt{15}} \,\hop{\bar{p}}{8} + \frac{1}{\sqrt{3}} \,\hop{\bar{p}}{9} \, ,
\nonumber\\
\hop{\delta}{6} &=& \frac{1}{\sqrt{3}} \,\hop{\bar{p}}{2} + \frac{1}{\sqrt{15}} \,\hop{\bar{p}}{3} + \frac{1}{\sqrt{35}} \,\hop{\bar{p}}{4}
+ \frac{1}{2}\sqrt{\frac{65}{7}} \,\hop{\bar{p}}{6} + \frac{3}{2} \,\hop{\bar{p}}{10}\, ,
\nonumber\\
\hop{\delta}{7} &=& \sqrt{\frac{7}{5}} \,\hop{\bar{p}}{7} + \frac{2}{\sqrt{15}} \,\hop{\bar{p}}{8} + \frac{2}{\sqrt{3}} \,\hop{\bar{p}}{9} \, ,
\nonumber\\
\hop{\delta}{8} &=& \sqrt{\frac{5}{3}} \,\hop{\bar{p}}{8} + \frac{2}{\sqrt{3}} \,\hop{\bar{p}}{9}  \, ,
\nonumber\\
\hop{\delta}{9} &=& \sqrt{3} \,\hop{\bar{p}}{9} \, ,
\nonumber\\
\hop{\delta}{10} &=& 2 \,\hop{\bar{p}}{10} \, .
\label{delta_3R}
\end{eqnarray}

Before proceeding to the discussion for general $N$-rail configurations, we would like to point out here that, as previously for canonical
clusters, the two-rail results \eref{mu_2R_lo}, \eref{nu_2R_lo} and the three-rail results \eref{mu_3R_lo},
\eref{nu_3R_lo} correspond to the minimum excess noises in the quadratures $\hop{q}{\mu}$ and $\hop{q}{\nu}$
when the nodes are uniformly squeezed. For example, for the two-rail results
\eref{mu_2R_lo} and \eref{nu_2R_lo}, one can have instead for $\hop{q}{\mu}$ and $\hop{q}{\nu}$
the general expressions
\begin{eqnarray}
\hat{q}_\mu &\equiv& \hat{q}_1 - (\eta_1\,\hop{p}{2}+\eta_2\,\hop{p}{3}) + \sqrt{2}\hat{q}_{\alpha_1}
\nonumber \\
&=& \hat{q}_\alpha
- \left[\eta_1\,(\hat{p}_2 - \hat{q}_1 - \hop{q}{4}) +\eta_2\,(\hat{p}_3 - \hat{q}_1 - \hop{q}{4}) \right] \, ,
\nonumber \\
\hat{q}_\nu &\equiv& \hat{q}_8 - (\eta_3\,\hop{p}{6}+\eta_4\,\hop{p}{7}) + \sqrt{2}\hat{q}_{\beta_1}
\nonumber \\
&=& \hat{q}_\beta
- \left[\eta_3\,(\hat{p}_6 - \hat{q}_5 - \hop{q}{8}) +\eta_4\,(\hat{p}_7 - \hat{q}_5 - \hop{q}{8}) \right] \, ,
\label{qq_2R_lo}
\end{eqnarray}
where $\eta_1+\eta_2=1$ and $\eta_3+\eta_4=1$. Despite the more complicated noise operators \eref{delta_2R},
for identically squeezed cluster nodes, one can again show that it is
the symmetrical arrangement $\eta_k=1/2$ for all $k=1\sim 4$ in \eref{mu_2R_lo} and \eref{nu_2R_lo}
that would minimize the excess noise in the teleported $\hop{q}{\mu}$ and $\hop{q}{\nu}$. Similarly, this is
also the case with the three-rail results in \eref{mu_3R_lo} and \eref{nu_3R_lo}. As we had noted
when studying multi-rail canonical clusters, optimization through the choice of coefficients such as in \eref{qq_2R_lo}
is the underlying mechanism for excess-noise reduction through multi-rail designs. We will come back to this
issue shortly in our discussion for general $N$-rail clusters in the following.

We are now in a position to extend the results above to general linear-optical cluster states with an
arbitrary $N$-rail design. Upon inspecting the results \eref{mu_L6_lo}, \eref{nu_L6_lo}, \eref{mu_2R_lo},
\eref{nu_2R_lo}, \eref{mu_3R_lo}, and \eref{nu_3R_lo}, for the CZ-gate teleportation in Fig.~\ref{fig:MR_cz}(d)
with an $N$-rail linear-optical cluster, one can obtain by induction the output quadratures
\begin{eqnarray}
\hat{q}_\mu &=& \hat{q}_\alpha - \frac{1}{N} \sum_{k=2}^{N+1}\hop{\delta}{k} \, ,
\nonumber\\
\hat{p}_\mu &=& \hat{p}_\alpha + \hop{q}{\beta} + \hop{\delta}{1} - \hop{\delta}{N+2} \, ,
\nonumber\\
\hat{q}_\nu &=& \hat{q}_\beta - \frac{1}{N} \sum_{k=N+4}^{2N+3}\hop{\delta}{k} \, ,
\nonumber\\
\hat{p}_\nu &=& \hat{p}_\beta + \hop{q}{\alpha} + \hop{\delta}{2N+4} - \hop{\delta}{N+3} \, .
\label{qp_NR_lo}
\end{eqnarray}
As in \eref{qp_NR}, here we have expressed the formulas in favor of the noise operators \eref{noise_lo}
both for compactness and for later convenience in calculating the quadrature correlators.
Although \eref{qp_NR_lo} is in close resemblance with \eref{qp_NR}, one must be alert to the more
complicated correlations among the noise operators $\hop{\delta}{k}$ here than those of canonical clusters.
Remarkably, for any linear-optical cluster with uniformly squeezed nodes, one can derive an
analytical expression for its noise correlators
(see Appendix \ref{sec:proof})
\begin{eqnarray}
\langle \hop{\delta}{k}\,\hop{\delta}{l}\rangle
= (M_{kl} + \delta_{kl}) \times \frac{e^{-2r}}{4} \, ,
\label{miracle}
\end{eqnarray}
where $r$ is the squeezing parameter for the nodes and
\begin{eqnarray}
M_{kl} \equiv
\left\{
\begin{array}{lr}
\mbox{(number of nodes connected to node $k$),} & \mbox{if $k=l$;} \\
\mbox{(number of nodes connected to both nodes $k$ and $l$),}& \mbox{if $k\neq l$.}
\end{array}
\right.
\label{Mkl}
\end{eqnarray}
This result will allow us to find an analytical formula for the entanglement in the
output modes $\mu$, $\nu$ of the teleported CZ-gate for general $N$-rail linear-optical
cluster. In addition, we can make use of \eref{miracle} to understand the noise-reduction
mechanism behind the multi-rail scheme for teleportation with linear-optical clusters.
That is, as with canonical clusters (see below \eref{q_gen_NR}),
in our expressions for $\hop{q}{\mu}$ and $\hop{q}{\nu}$ in \eref{qp_NR_lo},
the coefficient $\frac{-1}{N}$ in the sum over noise operators $\hop{\delta}{k}$ corresponds
to allocating the excess noise to each of the multi-rails with equal weight.
As shown in Appendix \ref{sec:proof}, this symmetric arrangement would minimize the excess noise
in the teleported $\hop{q}{\mu}$ and $\hop{q}{\nu}$.

Equipped with \eref{miracle}, we shall now examine the quality of the teleported CZ-gate through an
$N$-rail cluster by calculating the log-negativity for the output modes. As always, we shall assume
that we have independent coherent-state input modes $\alpha$, $\beta$, and that the cluster nodes
are uniformly squeezed with squeezing parameter $r$. With the help of \eref{miracle},
the quadrature correlators \eref{abc} can then be found readily through \eref{qp_NR_lo}, which yield
\begin{eqnarray}
a = \frac{(1+\frac{2N+1}{N}e^{-2r})}{4} \, , \quad
b = \frac{(2+3e^{-2r})}{4} \, , \quad
c = \frac{(1+e^{-2r})}{4} \, .
\label{abc_NR_lo}
\end{eqnarray}
As with their counterparts for canonical clusters, here $b$ and $c$ are again $N$-independent.
Using \eref{abc_NR_lo} in \eref{EN2}, one finds that the log-negativity for the output modes in
this case reads
\begin{eqnarray}
E_{N,NR'} = \max\left\{0,-\ln\!\left(\sqrt{(1+\frac{(2N+1)e^{-2r}}{N})(2+3e^{-2r})}-(1+e^{-2r})\right)\right\} \, .
\label{EN_NR_lo}
\end{eqnarray}

\begin{figure}
\includegraphics*[width=90mm]{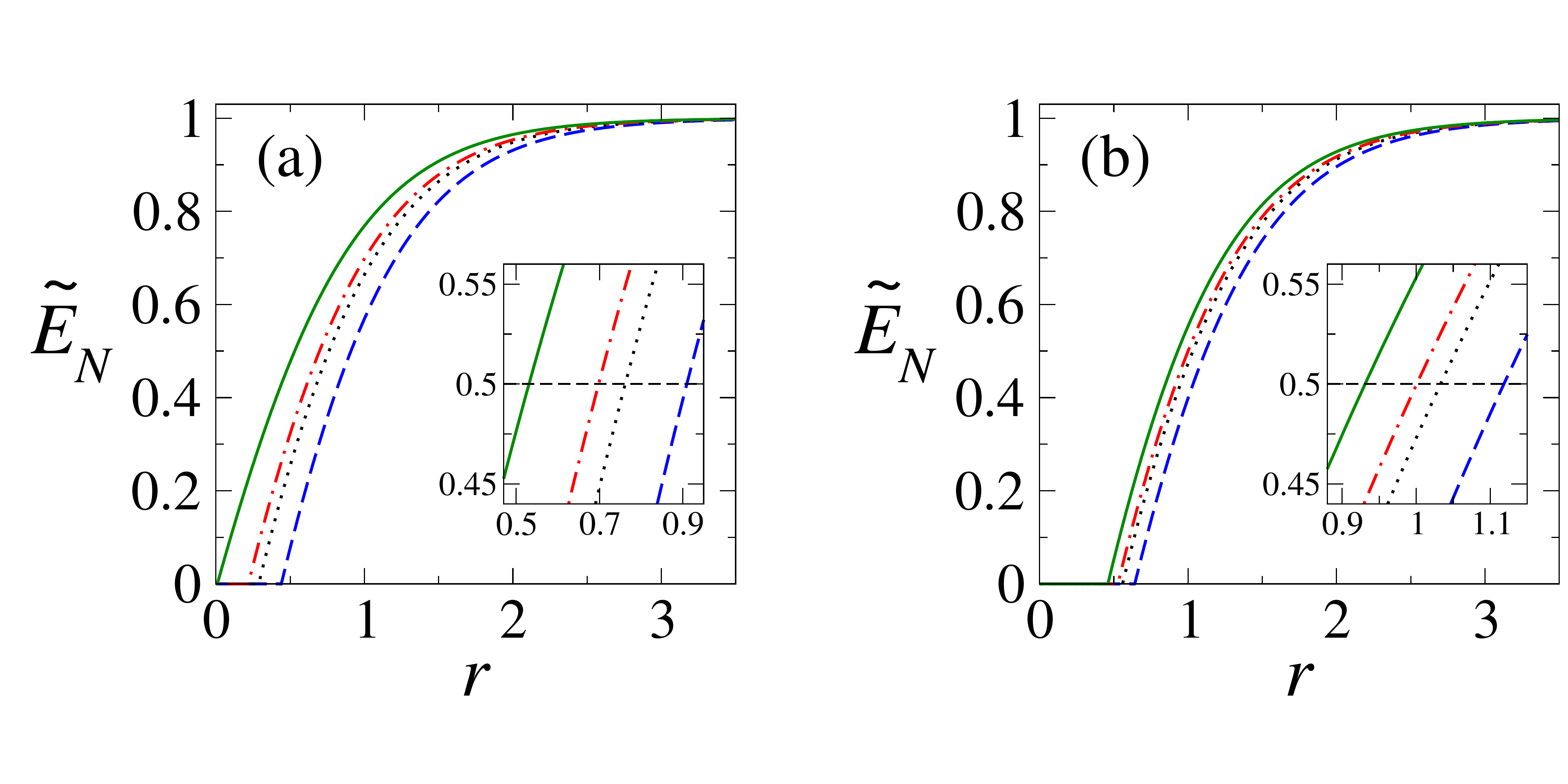}
\caption{The log-negativity $\tilde{E}_N$ for the output modes $\mu$, $\nu$ of the teleported
CZ-gate normalized with respect to the ideal (infinite squeezing) value
plotted as a function of the squeezing parameter $r$. Here the resource states are (a) canonical and (b)
linear-optical multi-rail clusters with the number of rails
$N=1$ (dashed lines), 2 (dotted lines), 3 (dot-dashed lines), and
100 (solid lines). In each panel, the inset shows the vicinity of the
squeezing level at which the log-negativity reaches 50$\%$ of the ideal
value. \label{fig:EN}}
\end{figure}

With the entanglement of the output modes available for CZ-gate
teleportation through canonical cluster states and linear-optical
cluster states, we are now ready for comparisons between them. We
plot in Fig.~\ref{fig:EN} the log-negativities \eref{EN_NR} and
\eref{EN_NR_lo} with respect to the squeezing parameter $r$ for
different numbers of rails $N$, where $E_N$ is normalized relative
to the ideal (infinite squeezing) value
$E_N(\infty)=-\ln(\sqrt{2}-1)\simeq 0.88$. We note that for the
linear-optical result \eref{EN_NR_lo}, the log-negativity approaches
the result \eref{EN_L4_lo} for a linear four-mode cluster in the
limit of $N\rightarrow \infty$, which remains to have regimes with
zero entanglement for $r\lesssim 0.45$ (visible from the plot for
$N=100$ in Fig.~\ref{fig:EN}(b)). In contrast, for canonical cluster
states, the log-negativity \eref{EN_NR} would be identical to the
corresponding linear four-mode result \eref{EN_L4} when $N=2$, and
for large $N$ the output modes can have non-zero entanglement even
at very low squeezing levels (e.g., see the plot for $N=100$ in
Fig.~\ref{fig:EN}(a)).

\begin{table}
\caption{Squeezing parameter $\rb$ for achieving 50$\%$ of the ideal
log-negativity with $N$-rail clusters} \label{tab:r}
\begin{ruledtabular}
\begin{tabular}{ lcccccc }
&\multicolumn{5}{c}{ $N$ } \\
\cline{2-6}
& 1 & 2 & 3 & 100 & $\infty$ \\
\hline
$\rb$ (can.)\footnote[1]{canonical clusters} & 0.91 & 0.76 & 0.70 & 0.53 & $0.52+{\rm O}(1/N)$ \\
$\rb$ (l.o.)\footnote[2]{linear-optical clusters} & 1.12 & 1.03 & 1.00 & 0.93 & $0.93+{\rm O}(1/N)$
\end{tabular}
\end{ruledtabular}
\end{table}

As a figure of merit for the CZ-gate teleportation, let us consider
the squeezing parameter $\rb$ that would enable the teleportation to
achieve one half of the ideal value for the log-negativity (see the
insets in Fig.~\ref{fig:EN}). Analytic expressions for $\rb$ can be
obtained by solving the equation
\begin{eqnarray}
\sqrt{ab}-c = \frac{\sqrt{\sqrt{2}-1}}{4} \, ,
\label{rb}
\end{eqnarray}
where the correlators $a$, $b$, $c$ are given by \eref{abc_NR} for
canonical clusters and \eref{abc_NR_lo} for linear-optical clusters.
We list in Table \ref{tab:r} the values of $\bar{r}$ for both types
of multi-rail clusters with selected number of rails $N$. It is seen
that for linear-optical clusters, $\rb$ starts with the single-rail
value 1.12 (corresponding to $-9.73$ dB) and, with the
implementation of multi-rail structures, reduces to 1.00 ($-8.69$
dB) for $N=3$ and to 0.93 ($-8.08$ dB) for $N=100$. Judging from the
reduction of $\rb$ achieved by increasing $N$ in the multi-rail, we
see that the improvement in the CZ-gate teleportation is rather
limited in this case. In the instance of the canonical clusters,
$\rb$ takes the value 0.91 ($-7.90$ dB) for single-rail and
decreases appreciably when multi-rail design is incorporated: $\rb$
becomes 0.70 ($-6.08$ dB) when $N=3$ and 0.53 ($-4.60$ dB) when
$N=100$. If this trend could persist for even larger values of $N$,
so that $\rb$ would tend to zero for sufficiently large $N$, one
would then be able to achieve ideal log-negativity for the CZ-gate
teleportation via multi-rail canonical clusters with vanishing
squeezing. Nonetheless, as one can show analytically, for both
linear-optical and canonical clusters $\rb$ would approach non-zero
steady values in the large $N$ limit in the manner of $1/N$ (see
Table \ref{tab:r}). Therefore, despite the impressive reduction of
$\rb$ for multi-rail canonical clusters, it is not possible to
reduce $\rb$ without bound towards zero. In other words, for either
class of resource states considered here, multi-rail noise reduction
is insufficient to enable ideal CZ-gate teleportation if the
resource state had not been sufficiently squeezed.

To understand the reason for the difference between the results for
the two classes of resource states, let us recall that the
entanglement between the output modes is determined solely from the
smaller symplectic eigenvalue $\lambda_-$ in \eref{symp} for the
partially transposed covariance matrix of the output modes, which,
in turn, is governed by the quadrature correlators \eref{abc}. From
the results \eref{abc_NR} and \eref{abc_NR_lo}, we find that for
both canonical and linear-optical clusters, the correlators $b$ and
$c$ for the teleported states are entirely independent of the number
of rails $N$. Therefore, the $N$-dependence of $\lambda_-$ (and
hence, of the log-negativity $E_N$) is completely due to the
correlator $a$. Since we have $\sqrt{ab} > c$ always for both
\eref{abc_NR} and \eref{abc_NR_lo}, according to \eref{symp} and
\eref{EN2}, the entanglement thus changes monotonically with $a$
when $b$ and $c$ stay fixed. In the case of canonical clusters, we
see from \eref{abc_NR} that the multi-rail design can eliminate the
excess noise in $a$ entirely in the large $N$ limit, so that
$\lambda_-$ can drop below the critical value $1/4$ for
entanglement. Thus, with increasing number of rails, the
log-negativity \eref{EN_NR} would increase steadily and tend to a
limit with non-vanishing entanglement for the whole range of
squeezing parameter $r$. In contrast, for linear-optical clusters,
the multi-rail noise reduction can bring down $a$ at best to the
linear four-mode expression \eref{abc_L4_lo} even with
$N\rightarrow\infty$. The entanglement \eref{EN_NR_lo} is thus
always bounded above by $E_{N,L4'}$ of \eref{EN_L4_lo}, which
vanishes for a finite range of squeezing levels. Likewise, for the
figure of merit $\rb$, one can also understand why the multi-rail
reduction works better for canonical clusters than for
linear-optical ones. This is because in the large $N$ limit, only
the correlator $b$ would be left with excess-noise term in the case
of canonical clusters, while all three correlators $a$, $b$, and $c$
would still have excess-noise terms for linear-optical clusters. In
fact, this is also why one can never succeed in reducing $\rb$ to
zero even with infinite $N$, since there are always excess-noise
terms left in the quadrature correlator(s). If one could design
cluster-type resource states for which the output mode quadrature
correlators would depend on some parameters that could erase {\em
all} excess-noise terms under appropriate limits, it would then be
possible to achieve perfect CZ-gate teleportation with resource
states of {\em arbitrarily low} squeezing.

\begin{figure}
\includegraphics*[width=90mm]{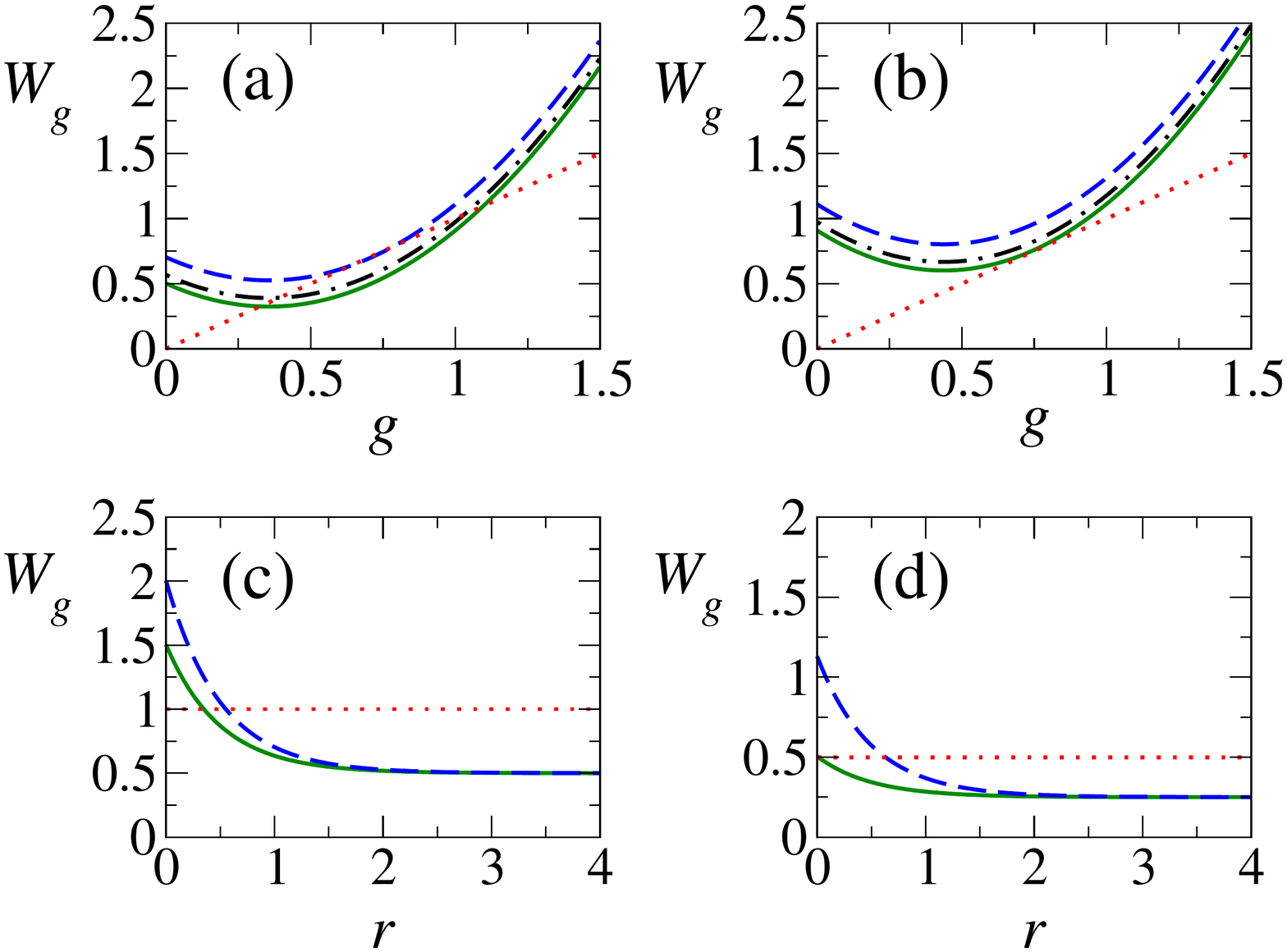}
\caption{Panels (a) and (b): $W_g$ as a function of
the signal gain $g$ at $-3.91$ dB ($r=0.45$) squeezing for teleportations with
(a) canonical and (b) linear-optical clusters with multi-rails $N=$ 1 (dashed lines),
3 (dot-dashed lines), and 100 (solid lines). Panels (c) and (d):
$W_g$ plotted versus the squeezing parameter $r$ at fixed gain values (c) $g=1.0$
and (d) $g=0.5$ for teleportations with canonical (solid lines) and
linear-optical (dashed lines) clusters with the number of rails $N=100$.
In each panel, the dotted line depicts the corresponding entanglement
bound.
\label{fig:Wg}}
\end{figure}

In order to verify our results experimentally, one can construct the
linear-optical clusters, as usual, by passing momentum-squeezed
vacuum modes through appropriate networks of beam splitters and
phase shifters \cite{vL07,Uk11a,Uk11b,Su13,Fu11}. The number of
optical elements required, nevertheless, would increase rapidly with
the number of multi-rails implemented. For the canonical clusters,
one may achieve the multi-rail structure by means of cluster shaping
over universal two-dimensional cluster states \cite{Mi10}. This is,
however, rather wasteful for the resource states and the number of
rails achievable would also be limited by the connectivity of the
nodes in the original cluster. Alternatively, one may resort to the
temporal-encoding based single-QND approach \cite{Me10}. By routing
and controlling the qumodes exquisitely, it is possible to fabricate
the canonical multi-rail cluster employing a single QND-gate (see
Appendix \ref{sec:1QND}).

For the entanglement detection, rather than the full covariance
matrix, experimentally it is favorable to check through the variance
sum of suitable quadrature combinations, such as \cite{Uk11b}
\begin{eqnarray}
W_g\equiv\langle[\Delta(g\hop{p}{\mu}-\hop{q}{\nu})]^2\rangle +
\langle[\Delta(g\hop{p}{\nu}-\hop{q}{\mu})]^2\rangle \, ,
\label{Wg}
\end{eqnarray}
where $g$ is a signal gain introduced to sharpen the entanglement
bound (see below). Measurement with $W_g<g$ indicates entanglement
between the output modes $\mu$ and $\nu$. The variance function
$W_g$ can be readily expressed in terms of the correlators
\eref{abc}, which yields $W_g=2(a+g^2b-2gc)$. Making use of our
results \eref{abc_NR} for canonical clusters, and \eref{abc_NR_lo}
for linear-optical clusters, we plot in Fig.~\ref{fig:Wg}(a) and (b)
the dependence of $W_g$ on the gain factor $g$ for canonical and
linear clusters with different multi-rail structures at the
squeezing level $r=0.45$ ($-3.91$ dB). We see that for canonical
clusters, the teleported modes become entangled with the
implementation of the multi-rails, while for linear-optical
clusters, the teleported modes remains to have no detectable
entanglement even with large number of multi-rails $N=100$. This is
consistent with our finding through log-negativities shown in
Fig.~\ref{fig:EN}.

It should be noted that the gain factor $g$ is crucial here for the
entanglement detection. As illustrated in Fig.~\ref{fig:Wg}(c) for
the canonical cluster with multi-rails of $N=100$, with unit gain
($g=1$), $W_g$ would fail to detect entanglement for $r\lesssim
0.35$, despite the presence of entanglement even at $r\simeq 0.01$
according to the log-negativity [see Fig.~\ref{fig:EN}(a)]. In
contrast, as depicted in Fig.~\ref{fig:Wg}(d), the choice of $g=0.5$
changes the situation entirely: $W_g$ would then be able to detect
entanglement even at very low squeezing levels. In fact, for the
optimal gain $g = (c+1/4)/b$ that minimizes $(W_g - g)$ (thus
providing the most stringent entanglement bound for a given
configuration), one can show easily that the condition $W_g<g$ for
entanglement would become exactly identical to the
symplectic-eigenvalue based criterion $\lambda_-=(\sqrt{ab}-c)<1/4$.

\section{\label{sec:concl}Conclusion and Discussions}
In summary, we have examined in this paper the teleportation of a
CZ-gate through measurement-based scheme in the CV regime, where two
classes of resource states have been considered: the canonical
cluster states fabricated through QND-gates and the linear-optical
cluster states from linear-optical networks. We study the
entanglement properties of the teleported output modes and examine
the performance of a multi-rail design that aims to reduce the
excess noise in the teleported gate. For both classes of resource
states, we find analytical expressions for the entanglement and
obtain its scaling with the number of rails in the multi-rail
design. It is found that the multi-rail design can help improve the
CZ-gate teleportation significantly when canonical clusters are
adopted. In the process of this analysis, we also discuss in detail
the noise-reduction mechanism underlying the multi-rail approach to
measurement-based teleportation. And the last but not the least, in
order to facilitate the analysis for teleportation with multi-rail
clusters, we have developed an operator-manipulation trick that can
help establish efficiently the measurement sequence and the
corrective operations for the CZ-gate teleportation.

Although the quadrature-manipulation trick has been applied in this work only to CZ-gate
teleportation, it is equally applicable to teleportation of other gates
in measurement-based quantum information processing. For instance, for single-qumode
gates that effect translation, rotation, and squeezing of a single qumode, one can apply the same trick
and obtain the corresponding measurement sequence and corrective operations in the measurement-base
teleportation \cite{Su15}.

It must be noted that the entanglement enhancement in the teleported
CZ-gate in our result should not be interpreted as a scheme for
building an ``improved" CZ-gate, which can be used, for instance, in
the single-QND approach to cluster construction \cite{Me10}.
Instead, the CZ-gate teleportation here is used as a task for
testing the performance of a given resource state in the
measurement-based scheme. Our result demonstrates that by
implementing multi-rail designs into the resource states, due to
suppression of the excess noise, it is possible to entangle two
input modes more effectively than the original single-rail clusters.
In the light of this result, therefore, it is then interesting to
enquire whether this multi-rail design could help lower the
squeezing threshold for fault-tolerant measurement-based CV quantum
computing when appropriate error-correcting code is incorporated
\cite{Me14}.

Finally, in closing we note that despite the lower quality of the teleported CZ-gate through linear-optical
cluster states in comparison with that via canonical cluster states in our results, it can be possible to
change this situation by exploiting an additional degree of freedom in constructing the $U$-matrix for the
linear-optical cluster. Since the geometric constraints resulting from \eref{geo_con} consist of inner products of
the $\vec{\alpha}_k$'s (i.e., the $G_{kl}$ of \eref{Gkl}), they are invariant under orthogonal
transformations. In other words, for any choice of the $\vec{\alpha}_k$'s one can always apply
orthogonal transformations over these vectors without violating the geometric constraints. This extra
degree of freedom thus provides means of optimization for constructing the $U$-matrix according
to the cost function one choose to consider \cite{Fer15}. Since the number of nodes can be quite large in our
calculation, such optimization can be challenging. Also, it is possible to improve the quality
of the teleported gates by introducing appropriate gain factors in the feedback signals \cite{Zh06,Fu11}.
We plan to investigate these issues of optimization in a separate work.

We thank Prof. Dian-Jiun Han for useful discussions, and Dr.~Yi-Chan
Lee for valuable suggestions and help in parts of this work. This
research is supported by the MOST of Taiwan through grant MOST
104-2112-M-194-006. It is also supported partly by the Center for
Theoretical Sciences, Taiwan.

\appendix
\section{\label{sec:FTc}CZ-gate teleportation using Fourier transformed cluster states}
In this appendix we demonstrate that for the CZ-gate teleportation of Fig.~\ref{fig:L4-2}
with a canonical linear four-mode cluster, one can perform the quadrature
manipulations following the scheme of category (d), rather than those of categories (b) and (c),
and arrive at results equivalent to those of Sec.~\ref{sec:form}. For this purpose, it
is necessary to apply Fourier transforms over the cluster modes prior to their coupling with
the input modes.
The quadrature operators for the Fourier-transformed cluster nodes $k=1\sim 4$ read
\begin{eqnarray}
\left(
  \begin{array}{c}
     \hat{q}_k'\\
     \hat{p}_k'\\
  \end{array}
\right)
=
\left(
  \begin{array}{c}
    -\hat{p}_k\\
    +\hat{q}_k\\
  \end{array}
\right)
=
\left(
  \begin{array}{c}
     -\hop{\bar{p}}{k}-\sum_{l\in N_k} \hop{\bar{q}}{l}\\
     \hop{\bar{q}}{k}\\
  \end{array}
\right) \, ,
\label{FT_nodes}
\end{eqnarray}
where, as always, $\hop{\bar{q}}{k}$ and $\hop{\bar{p}}{k}$ are quadrature operators for the
offline-squeezed initial cluster mode $k$, and we have used \eref{qp_QND} in reaching the last expression.
It follows immediately from \eref{FT_nodes} that, instead of
\eref{noise_can}, the excess-noise operators (or nullifiers) here become
\begin{eqnarray}
\hat{\delta}_k \equiv \hat{q}_k' + \sum_{l\in N_k} \hat{p}_l' = - \hat{\bar{p}}_k \, .
\label{noise_FT}
\end{eqnarray}
QND-couplings between the input modes and the cluster nodes thus yield (cf.~\eref{in_coup2})
\begin{eqnarray}
\left(
  \begin{array}{c}
     \hat{q}_\rho''\\
     \hat{p}_\rho''\\
  \end{array}
\right)
=
\left(
  \begin{array}{c}
    \hat{q}_\rho\\
    \hat{p}_\rho + \hat{q}_k'\\
  \end{array}
\right)
\quad \mbox{and} \quad
\left(
  \begin{array}{c}
     \hat{q}_k''\\
     \hat{p}_k''\\
  \end{array}
\right)
=
\left(
  \begin{array}{c}
     \hat{q}_k'\\
     \hat{p}_k' + \hat{q}_\rho\\
  \end{array}
\right) \, ,
\label{in_coup_FT}
\end{eqnarray}
where $(\rho,k)=(\alpha,2)$ and $(\beta,3)$, and we have denoted all resultant modes with
double-primed notations for clarity.

In anticipation of the mapping
\begin{eqnarray}
\left(
  \begin{array}{c}
    \hat{q}_1'\\
    \hat{p}_1'\\
  \end{array}
\right)
\rightarrow
\left(
  \begin{array}{c}
    \hat{q}_\alpha \\
    \hat{p}_\alpha + \hat{q}_\beta\\
  \end{array}
\right)
\quad \mbox{and} \quad
\left(
  \begin{array}{c}
    \hat{q}_4'\\
    \hat{p}_4'\\
  \end{array}
\right)
\rightarrow
\left(
  \begin{array}{c}
    \hat{q}_\beta \\
    \hat{p}_\beta + \hat{q}_\alpha\\
  \end{array}
\right)
\label{cz_FT_goal}
\end{eqnarray}
subject to proper corrective operations, one can write employing the
$\hop{p}{2}''$ entry of \eref{in_coup_FT}
\begin{eqnarray}
\hat{q}_1' &=& \hat{q}_1' - (\hat{p}_2'' - \hop{p}{2}' - \hat{q}_\alpha)
\nonumber \\
&=& \hat{q}_\alpha - \hat{p}_2'' + (\hat{q}_1' + \hat{p}_2') \,
\label{q1_FT}
\end{eqnarray}
with $(\hat{q}_1' + \hat{p}_2')$ in the last line being the nullifier $\hop{\delta}{1}$
of \eref{noise_FT}. The last expression above immediately suggests that
\begin{eqnarray}
\hat{q}_\mu &\equiv& \hat{q}_1' + \hat{p}_2''
\nonumber \\
&=&  \hat{q}_\alpha + (\hat{q}_1' + \hat{p}_2') \, .
\label{qmu_FT}
\end{eqnarray}
Similarly, in view of \eref{cz_FT_goal}, we can write with the help of \eref{in_coup_FT}
\begin{eqnarray}
\hat{p}_1' &=& \hat{p}_1' - (\hat{p}_\alpha'' - \hop{p}{\alpha} - \hat{q}_2') - (\hat{p}_3'' - \hat{p}_3' - \hat{q}_\beta )
\nonumber \\
&=& \hat{p}_\alpha + \hat{q}_\beta - \hat{p}_\alpha'' - \hat{p}_3'' + (\hat{q}_2' + \hat{p}_1' + \hat{p}_3') \, .
\label{p1_FT}
\end{eqnarray}
It thus follows that
\begin{eqnarray}
\hat{p}_\mu &\equiv& \hat{p}_1' + \hat{p}_\alpha'' + \hat{p}_3''
\nonumber \\
&=&  \hat{p}_\alpha + \hat{q}_\beta + (\hat{q}_2' + \hat{p}_1' + \hat{p}_3') \,.
\label{pmu_FT}
\end{eqnarray}
As before, by virtue of the symmetry among the modes, for the output mode $\nu$
one can obtain from \eref{qmu_FT} and \eref{pmu_FT} by changing the indices suitably
\begin{eqnarray}
\hat{q}_\nu &\equiv& \hat{q}_4' + \hat{p}_3''
\nonumber \\
&=&  \hat{q}_\beta + (\hat{q}_4' + \hat{p}_3') \, ,
\nonumber\\
\hat{p}_\nu &\equiv& \hat{p}_4' + \hat{p}_\beta'' + \hat{p}_2''
\nonumber \\
&=&  \hat{p}_\beta + \hat{q}_\alpha + (\hat{q}_3' + \hat{p}_2' + \hat{p}_4') \, .
\label{nu_FT}
\end{eqnarray}
Comparing the second line of each expression for the output
quadratures above with its counterpart in \eref{qmu_L4-2},
\eref{pmu_L4-2}, and \eref{nu_L4-2}, we see that the output modes
here differ from those earlier only in sign changes in the
excess-noise terms. As far as the entanglement of the output modes
is concerned, which depends only on
$\langle\hop{\delta}{k}^2\rangle$, the results above are therefore
entirely equivalent to those found in Sec.~\ref{sec:form}.

\section{\label{sec:GU}Construction of the $U$-matrices for linear-optical cluster states}
We explain in this appendix  the procedure for constructing the unitary matrix $U$ in \eref{U_transf} which
represents the effects of a linear-optical network that implements the desired cluster
correlations for a linear-optical cluster state. The results of these calculations related to
the linear-optical cluster states considered in the text will also be supplied in the following.

In constructing the $U$-matrix for a given cluster geometry, one starts by solving from \eref{geo_con}
the geometric constraints over the row vectors $\vec{\alpha}_k$, which consist of
definite values for the inner products $G_{kl}\equiv\vec{\alpha}_k\,\vec{\alpha}_l^{\,T}$.
For a cluster with $M$ nodes, there are $M(M+1)/2$ such constraints that can be solved from \eref{geo_con} and
the result can be conveniently summarized in terms of an $M\times M$ real
symmetric matrix $G$, whose $k,l$ element is given by $G_{kl}$. For instance, for the
linear four-mode cluster in Figs.~\ref{fig:L4-1} and \ref{fig:L4-2}, we find
\begin{eqnarray}
G_{L4}=
\left(
  \begin{array}{cccc}
    \frac{3}{5} & 0 & \frac{-1}{5} & 0 \\
    0 & \frac{2}{5} & 0 & \frac{-1}{5} \\
    \frac{-1}{5} & 0 & \frac{2}{5} & 0 \\
    0 & \frac{-1}{5} & 0 & \frac{3}{5} \\
  \end{array}
\right) \, ,
\label{G_L4}
\end{eqnarray}
where the subscript $L4$ stands for ``linear four-mode". With the geometric constraints,
it is then straightforward to construct the vectors $\vec{\alpha}_k$ accordingly.
For this task, it is advisable to start from the ``symmetric center" of the cluster. In the present
case of the linear four-mode cluster, we start from nodes 2 and 3 in keeping with the geometric constraints
$(G_{L4})_{22}=(G_{L4})_{33}=2/5$ and that $(G_{L4})_{23}=0$ from \eref{G_L4} and take
\begin{eqnarray}
\vec{\alpha}_2 &=& \left(
                   \begin{array}{cccc}
                     0 & \sqrt{\frac{2}{5}} & 0 & 0 \\
                   \end{array}
                 \right) \, ,
\nonumber \\
\vec{\alpha}_3 &=& \left(
                   \begin{array}{cccc}
                     0 & 0 & \sqrt{\frac{2}{5}} & 0 \\
                   \end{array}
                 \right) \, .
\label{a_L4-1}
\end{eqnarray}
One can next choose to construct $\vec{\alpha}_1$ taking into account the geometric
constraints $(G_{L4})_{1k}$ with $k=1\sim 3$, leaving out $(G_{L4})_{14}$ for later
when constructing $\vec{\alpha}_4$. We get accordingly
\begin{eqnarray}
\vec{\alpha}_1 &=& \left(
                   \begin{array}{cccc}
                     \frac{1}{\sqrt{2}} & 0 & \frac{-1}{\sqrt{10}} & 0 \\
                   \end{array}
                 \right) \, .
\label{a_L4-2}
\end{eqnarray}
Taking into account the remaining constraints for $\vec{\alpha}_4$,
one can find immediately
\begin{eqnarray}
\vec{\alpha}_4 &=& \left(
                   \begin{array}{cccc}
                     0 & \frac{-1}{\sqrt{10}} & 0 & \frac{1}{\sqrt{2}} \\
                   \end{array}
                 \right) \, .
\label{a_L4-3}
\end{eqnarray}
With the $\vec{\alpha}_k$'s available, it is then straightforward to construct the
$k$-th row of the matrix $U$ following \eref{r_vec}, and thus the $U$-matrix.
Using \eref{a_L4-1}--\eref{a_L4-3}, we obtain for
the linear four-mode cluster
\begin{eqnarray}
U_{L4}=
\left(
  \begin{array}{cccc}
    \frac{1}{\sqrt{2}} & \frac{2i}{\sqrt{10}} & \frac{-1}{\sqrt{10}} & 0 \\
    \frac{i}{\sqrt{2}} & \frac{2}{\sqrt{10}} & \frac{i}{\sqrt{10}} & 0 \\
    0 & \frac{i}{\sqrt{10}} & \frac{2}{\sqrt{10}} & \frac{i}{\sqrt{2}} \\
    0 & \frac{-1}{\sqrt{10}} & \frac{2i}{\sqrt{10}} & \frac{1}{\sqrt{2}} \\
  \end{array}
\right) \,.
\label{U_L4}
\end{eqnarray}

For larger clusters, the procedure proceed similarly to that
presented above, although with greater complexity due to the
increased number of nodes and geometric constraints involved.
We summarize here our results
for the linear-optical cluster states that are considered in
Sec.~\ref{sec:lo}. For the linear six-mode cluster of
Fig.~\ref{fig:MR_cz}(a), we find the matrix for geometric
constraints
\begin{eqnarray}
G_{L6}=
\left(
  \begin{array}{cccccc}
    \frac{8}{13} & 0 & \frac{-3}{13} & 0 & \frac{1}{13} & 0 \\
    0 &  \frac{5}{13} & 0 & \frac{-2}{13} & 0 & \frac{1}{13} \\
    \frac{-3}{13} & 0 &  \frac{6}{13} & 0 & \frac{-2}{13} & 0 \\
    0 & \frac{-2}{13} & 0 &  \frac{6}{13} & 0 & \frac{-3}{13} \\
    \frac{1}{13} & 0 & \frac{-2}{13} & 0 &  \frac{5}{13} & 0 \\
    0 & \frac{1}{13} & 0 & \frac{-3}{13} & 0 &  \frac{8}{13} \\
  \end{array}
\right) \, .
\label{G_L6}
\end{eqnarray}
Constructing the vectors $\vec{\alpha}_k$ accordingly, we obtain the $U$-matrix
\begin{eqnarray}
U_{L6}=
\left(
  \begin{array}{cccccc}
    \frac{1}{\sqrt{2}} & \frac{i}{\sqrt{3}} & -\sqrt{\frac{3}{26}} & -\sqrt{\frac{2}{39}}i & 0 & 0 \\
    \frac{i}{\sqrt{2}} & \frac{1}{\sqrt{3}} & \sqrt{\frac{3}{26}}i & -\sqrt{\frac{2}{39}} & 0 & 0 \\
    0 & \frac{i}{\sqrt{3}} & \sqrt{\frac{6}{13}} &  2\sqrt{\frac{2}{39}}i & 0 & 0 \\
    0 & 0 & 2\sqrt{\frac{2}{39}}i & \sqrt{\frac{6}{13}} & \frac{i}{\sqrt{3}} & 0 \\
    0 & 0 & -\sqrt{\frac{2}{39}} & \sqrt{\frac{3}{26}}i & \frac{1}{\sqrt{3}} & \frac{i}{\sqrt{2}} \\
    0 & 0 & -\sqrt{\frac{2}{39}}i & -\sqrt{\frac{3}{26}} & \frac{i}{\sqrt{3}} & \frac{1}{\sqrt{2}} \\
  \end{array}
\right) \, .
\label{U_L6}
\end{eqnarray}
In the case of the two-rail cluster of Fig.~\ref{fig:MR_cz}(b), we find the geometric constraints
\begin{eqnarray}
G_{2R}=
\left(
  \begin{array}{cccccccc}
    \frac{9}{17} & 0 & 0 & -\frac{5}{17} & 0 & \frac{1}{17} & \frac{1}{17} & 0 \\
    0 & \frac{21}{34} & -\frac{13}{34} & 0 & -\frac{3}{34} & 0 & 0 & \frac{1}{17} \\
    0 & -\frac{13}{34} & \frac{21}{34} & 0 & -\frac{3}{34} & 0 & 0 & \frac{1}{17} \\
    -\frac{5}{17} & 0 & 0 & \frac{15}{34} & 0 & -\frac{3}{34} & -\frac{3}{34} & 0 \\
    0 & -\frac{3}{34} & -\frac{3}{34} & 0 & \frac{15}{34} & 0 & 0 & -\frac{5}{17} \\
    \frac{1}{17} & 0 & 0 & -\frac{3}{34} & 0 & \frac{21}{34} & -\frac{13}{34} & 0 \\
    \frac{1}{17} & 0 & 0 & -\frac{3}{34} & 0 & -\frac{13}{34} & \frac{21}{34} & 0 \\
    0 & \frac{1}{17} & \frac{1}{17} & 0 & -\frac{5}{17} & 0 & 0 & \frac{9}{17} \\
  \end{array}
\right) \, ,
\label{G_2R}
\end{eqnarray}
which allows us to construct the corresponding $U$-matrix
\begin{eqnarray}
U_{2R}=
\left(
  \begin{array}{cccccccc}
    \frac{1}{\sqrt{3}} & \frac{i}{\sqrt{3}} & \frac{i}{\sqrt{15}} & -\sqrt{\frac{10}{51}} & -\sqrt{\frac{6}{85}}i & 0 & 0 & 0 \\
    \frac{i}{\sqrt{3}} & \frac{1}{\sqrt{3}} & \frac{-2}{\sqrt{15}} & \sqrt{\frac{5}{102}}i & -\sqrt{\frac{3}{170}} & 0 & 0 & 0 \\
    \frac{i}{\sqrt{3}} & 0 & \sqrt{\frac{3}{5}} & \sqrt{\frac{5}{102}}i & -\sqrt{\frac{3}{170}} & 0 & 0 & 0 \\
    0 & \frac{i}{\sqrt{3}} & \frac{i}{\sqrt{15}} & \sqrt{\frac{15}{34}} & \sqrt{\frac{27}{170}}i & 0 & 0 & 0 \\
    0 & 0 & 0 & \sqrt{\frac{27}{170}}i & \sqrt{\frac{15}{34}} & \frac{i}{\sqrt{15}} & \frac{i}{\sqrt{3}} & 0 \\
    0 & 0 & 0 & -\sqrt{\frac{3}{170}} & \sqrt{\frac{5}{102}}i & \sqrt{\frac{3}{5}} & 0 & \frac{i}{\sqrt{3}} \\
    0 & 0 & 0 & -\sqrt{\frac{3}{170}} & \sqrt{\frac{5}{102}}i & \frac{-2}{\sqrt{15}} & \frac{1}{\sqrt{3}} & \frac{i}{\sqrt{3}} \\
    0 & 0 & 0 & -\sqrt{\frac{6}{85}}i & -\sqrt{\frac{10}{51}} & \frac{i}{\sqrt{15}} & \frac{i}{\sqrt{3}} & \frac{1}{\sqrt{3}} \\
  \end{array}
\right)
 \, .
\label{U_2R}
\end{eqnarray}
For the three-rail cluster in Fig.~\ref{fig:MR_cz}(c), we find
\begin{eqnarray}
G_{3R}=
\left(
  \begin{array}{cccccccccc}
    \frac{32}{65} & 0 & 0 & 0 & -\frac{21}{65} & 0 & \frac{3}{65} & \frac{3}{65} & \frac{3}{65} & 0 \\
    0 & \frac{47}{65} & -\frac{18}{65} & -\frac{18}{65} & 0 & -\frac{4}{65} & 0 & 0 & 0 & \frac{3}{65} \\
    0 & -\frac{18}{65} & \frac{47}{65} & -\frac{18}{65} & 0 & -\frac{4}{65} & 0 & 0 & 0 & \frac{3}{65} \\
    0 & -\frac{18}{65} & -\frac{18}{65} & \frac{47}{65} & 0 & -\frac{4}{65} & 0 & 0 & 0 & \frac{3}{65} \\
    -\frac{21}{65} & 0 & 0 & 0 & \frac{28}{65} & 0 & -\frac{4}{65} & -\frac{4}{65} & -\frac{4}{65} & 0 \\
    0 & -\frac{4}{65} & -\frac{4}{65} & -\frac{4}{65} & 0 & \frac{28}{65} & 0 & 0 & 0 & -\frac{21}{65} \\
    \frac{3}{65} & 0 & 0 & 0 & -\frac{4}{65} & 0 & \frac{47}{65} & -\frac{18}{65} & -\frac{18}{65} & 0 \\
    \frac{3}{65} & 0 & 0 & 0 & -\frac{4}{65} & 0 & -\frac{18}{65} & \frac{47}{65} & -\frac{18}{65} & 0 \\
    \frac{3}{65} & 0 & 0 & 0 & -\frac{4}{65} & 0 & -\frac{18}{65} & -\frac{18}{65} & \frac{47}{65} & 0 \\
    0 & \frac{3}{65} & \frac{3}{65} & \frac{3}{65} & 0 & -\frac{21}{65} & 0 & 0 & 0 & \frac{32}{65} \\
  \end{array}
\right) \, ,
\label{G_3R}
\end{eqnarray}
and the $U$-matrix can be constructed in the way described above. We get
\begin{eqnarray}
U_{3R}=
\left(
  \begin{array}{cccccccccc}
    \frac{1}{2} & \frac{i}{\sqrt{3}} & \frac{i}{\sqrt{15}} & \frac{i}{\sqrt{35}} & -\frac{3}{2}\sqrt{\frac{7}{65}} & \frac{-6i}{\sqrt{455}} & 0 & 0 & 0 & 0 \\
    \frac{i}{2} & \frac{1}{\sqrt{3}} & \frac{-2}{\sqrt{15}} & \frac{-2}{\sqrt{35}} & \frac{i}{2}\sqrt{\frac{7}{65}} & \frac{-2}{\sqrt{455}} & 0 & 0 & 0 & 0 \\
    \frac{i}{2} & 0 & \sqrt{\frac{3}{5}} & \frac{-2}{\sqrt{35}} & \frac{i}{2}\sqrt{\frac{7}{65}} & \frac{-2}{\sqrt{455}} & 0 & 0 & 0 & 0 \\
    \frac{i}{2} & 0 & 0 & \sqrt{\frac{5}{7}} & \frac{i}{2}\sqrt{\frac{7}{65}} & \frac{-2}{\sqrt{455}} & 0 & 0 & 0 & 0 \\
    0 & \frac{i}{\sqrt{3}} & \frac{i}{\sqrt{15}} & \frac{i}{\sqrt{35}} & 2 \sqrt{\frac{7}{65}} & \frac{8i}{\sqrt{455}} & 0 & 0 & 0 & 0 \\
    0 & 0 & 0 & 0 & \frac{8i}{\sqrt{455}} & 2 \sqrt{\frac{7}{65}} & \frac{i}{\sqrt{35}} & \frac{i}{\sqrt{15}} & \frac{i}{\sqrt{3}} & 0 \\
    0 & 0 & 0 & 0 & \frac{-2}{\sqrt{455}} & \frac{i}{2}\sqrt{\frac{7}{65}} & \sqrt{\frac{5}{7}} & 0 & 0 & \frac{i}{2} \\
    0 & 0 & 0 & 0 & \frac{-2}{\sqrt{455}} & \frac{i}{2}\sqrt{\frac{7}{65}} & \frac{-2}{\sqrt{35}} & \sqrt{\frac{3}{5}} & 0 & \frac{i}{2} \\
    0 & 0 & 0 & 0 & \frac{-2}{\sqrt{455}} & \frac{i}{2}\sqrt{\frac{7}{65}} & \frac{-2}{\sqrt{35}} & \frac{-2}{\sqrt{15}} & \frac{1}{\sqrt{3}} & \frac{i}{2} \\
    0 & 0 & 0 & 0  & \frac{-6i}{\sqrt{455}} & -\frac{3}{2}\sqrt{\frac{7}{65}} & \frac{i}{\sqrt{35}} & \frac{i}{\sqrt{15}} & \frac{i}{\sqrt{3}} & \frac{1}{2} \\
  \end{array}
\right) \, .
\label{U_3R}
\end{eqnarray}

It is the results \eref{U_L4}, \eref{U_L6}, \eref{U_2R}, and \eref{U_3R} for the $U$-matrices that we adopt in
our calculations for the linear-optical clusters in Sec.~\ref{sec:lo}, such as in finding the
excess-noise operators in \eref{delta_L4}, \eref{delta_L6}, \eref{delta_2R}, and \eref{delta_3R}.

\section{\label{sec:proof}Derivation for Eq.~\eref{miracle} and the noise-reduction mechanism for multi-rail linear-optical clusters}
Here we derive the identity \eref{miracle} for the noise correlators that is indispensable in calculating the
quadrature correlators \eref{abc} for linear-optical clusters. Based on \eref{miracle}, we will also provide
analysis for the noise reduction through multi-rail designs in measurement-based teleportation with linear-optical clusters.

To begin with, let us introduce a simplified notation by writing
$\vec{\beta}_k\equiv\sum_{l\in N_k} \vec{\alpha}_l$ in \eref{r_vec},
so that we have
\begin{eqnarray}
\vec{u}_k = \vec{\alpha}_k + i \vec{\beta}_k \, .
\label{r_vec2}
\end{eqnarray}
The orthonormality condition of $\vec{u}_k$ (or unitarity of $U$)
\eref{geo_con} can then be put in a more succinct form
\begin{eqnarray}
\sum_m \left(\alpha_{k,m}\,\alpha_{l,m} + \beta_{k,m}\,\beta_{l,m} \right) &=& \delta_{kl} \, ,
\nonumber\\
\sum_m \left(\alpha_{k,m}\,\beta_{l,m} - \alpha_{l,m}\,\beta_{k,m} \right) &=& 0 \, ,
\label{unit_ab}
\end{eqnarray}
where $\alpha_{k,m}$ denotes the $m$-th component of
$\vec{\alpha}_k$ and similarly for $\beta_{k,m}$. For the unitary
transformation \eref{U_transf} induced by a linear-optical network,
if we use the notation of \eref{r_vec2} and write in favor of the
quadrature operators, we would get
\begin{eqnarray}
\hop{q}{k} &=& \sum_l \left(\alpha_{k,l}\,\hop{\bar{q}}{l} - \beta_{k,l}\,\hop{\bar{p}}{l} \right) \, ,
\nonumber\\
\hop{p}{k} &=& \sum_l \left(\alpha_{k,l}\,\hop{\bar{p}}{l} + \beta_{k,l}\,\hop{\bar{q}}{l} \right) \, .
\label{qp_k}
\end{eqnarray}
It then follows that the noise operators are here
\begin{eqnarray}
\hop{\delta}{k} = \sum_m \left(\alpha_{k,m} + \sum_{n\in N_k} \beta_{n,m} \right) \hop{\bar{p}}{m} \, ,
\label{dk_ab}
\end{eqnarray}
which is \eref{noise_lo} expressed in the notation of \eref{r_vec2}. Since the initial
cluster modes are uncorrelated, it follows that
$\left\langle \hop{\bar{p}}{m} \,\hop{\bar{p}}{m'}\right\rangle
= \left\langle \hop{\bar{p}}{m}^2\right\rangle \delta_{mm'}$. The noise correlators
can thus be written
\begin{eqnarray}
\left\langle\hop{\delta}{k}\,\hop{\delta}{l}\right\rangle
= \sum_{m} \left[\left(\alpha_{k,m} + \sum_{n\in N_k} \beta_{n,m} \right) \,
\left(\alpha_{l,m} + \sum_{n'\in N_l} \beta_{n',m} \right) \,
\left\langle \hop{\bar{p}}{m}^2\right\rangle\right] \, .
\label{dkdk}
\end{eqnarray}

In the case when all initial cluster modes are momentum-squeezed
vacuum states with the same squeezing parameter $r$, we have
$\left\langle \hop{\bar{p}}{m}^2\right\rangle = e^{-2r}/4$ for every
mode $m$. One can then deal with the summations in \eref{dkdk}
making use of the unitarity of $U$ through \eref{unit_ab}. Factoring
out the constant $\langle \hop{\bar{p}}{m}^2\rangle$ in \eref{dkdk},
we spell out the summations there and arrive at
\begin{eqnarray}
\sum_{m} \left(\alpha_{k,m}\,\alpha_{l,m} + \alpha_{k,m} \sum_{n'\in N_l} \beta_{n',m}
+ \alpha_{l,m}\,\sum_{n\in N_k} \beta_{n,m} + \sum_{n\in N_k} \sum_{n'\in N_l} \beta_{n,m}\,\beta_{n',m} \right) \, .
\label{dkdk_sum1}
\end{eqnarray}
For the second term in the equation above, by exchanging the order of summations, we can write with the aid of \eref{unit_ab}
\begin{eqnarray}
\sum_{n'\in N_l} \left(\sum_m \alpha_{k,m} \beta_{n',m}\right)
= \sum_{n'\in N_l} \left(\sum_m \alpha_{n',m} \beta_{k,m}\right) \, .
\label{dkdk_sum2}
\end{eqnarray}
Exchanging again the order of summations in the last expression, we can arrive at
\begin{eqnarray}
\sum_m \beta_{k,m} \left( \sum_{n'\in N_l} \alpha_{n',m} \right) = \sum_m \beta_{k,m}\,\beta_{l,m}\, .
\label{dkdk_sum3}
\end{eqnarray}
For the third term in \eref{dkdk_sum1}, it is just the second term
there with $k$ and $l$ being exchanged. The result is thus the same
as that of \eref{dkdk_sum3}. For the fourth term in
\eref{dkdk_sum1}, again exchanging the order of summations there and
using \eref{unit_ab}, we can write
\begin{eqnarray}
\sum_{n\in N_k} \sum_{n'\in N_l} \left( \sum_m \beta_{n,m}\,\beta_{n',m} \right)
= \sum_{n\in N_k} \sum_{n'\in N_l} \left( \delta_{nn'} - \sum_m \alpha_{n,m}\,\alpha_{n',m} \right) \, .
\label{dkdk_sum4}
\end{eqnarray}
Exchanging again the order of summations in the second term above, we get
\begin{eqnarray}
\sum_{n\in N_k} \sum_{n'\in N_l} \delta_{nn'} - \sum_m \left(\sum_{n\in N_k} \alpha_{n,m}\right)\left(\sum_{n'\in N_l}\alpha_{n',m} \right)
= \sum_{n\in N_k} \sum_{n'\in N_l} \delta_{nn'} - \sum_m \beta_{k,m}\,\beta_{l,m} \, .
\label{dkdk_sum5}
\end{eqnarray}
Using the results \eref{dkdk_sum3} and \eref{dkdk_sum5} for the respective terms in \eref{dkdk_sum1}, we thus find
that the summations there become
\begin{eqnarray}
\sum_{n\in N_k} \left(\sum_{n'\in N_l}\delta_{nn'}\right)
+ \sum_m \left(\alpha_{k,m}\,\alpha_{l,m} + \beta_{k,m}\,\beta_{l,m} \right)
= M_{kl} + \delta_{kl} \, ,
\label{dkdk_sum_ans}
\end{eqnarray}
where we have used the first equation of \eref{unit_ab} in dealing
with the second summation on the left-hand side. For the first
summation in \eref{dkdk_sum_ans}, it yields exactly the $M_{kl}$
defined in \eref{Mkl}. Utilizing the result \eref{dkdk_sum_ans} in
\eref{dkdk}, we arrive finally at the identity \eref{miracle}.

In the light of the result \eref{miracle}, we will now demonstrate how noise reduction in the multi-rail
teleportation can be optimized through a symmetric arrangement among the rails of a linear-optical cluster.
In other words, we will show that the coefficient $\frac{-1}{N}$ in the expressions for $\hop{q}{\mu}$ and $\hop{q}{\nu}$
in \eref{qp_NR_lo} serves to minimize the excess noise in these output quadratures. Similar to \eref{q_gen_NR}
for canonical clusters, for CZ-gate teleportation through an $N$-rail linear-optical cluster, one can have
$\hop{q}{\mu}$ and $\hop{q}{\nu}$ in the generic expressions
\begin{eqnarray}
\hat{q}_\sigma = \hat{q}_\rho - \sum_k \eta_k\,\hop{\delta}{k} \, ,
\label{q_gen_NR_lo}
\end{eqnarray}
where $(\sigma,\rho)=(\mu,\alpha)$ and $(\nu,\beta)$, and the
summation over $k$ covers the same range as in \eref{qp_NR_lo}. Note
that here we have kept the minus sign in front of the summation for
consistency with its two-rail counterpart \eref{qq_2R_lo}. As
before, the coefficients $\eta_k$ in \eref{q_gen_NR_lo} must satisfy
the condition $\sum_k \eta_k =1$. Now that the noise operators
$\hop{\delta}{k}$ for linear-optical clusters are no longer
independent from each other, the excess noise in the teleported
$\hop{q}{\sigma}$ becomes here
$\langle(\hop{q}{\sigma}-\hop{q}{\rho})^2\rangle = \sum_{k,l}
\eta_k\eta_l\,\langle\hop{\delta}{k}\,\hop{\delta}{l}\rangle$. For a
uniformly squeezed linear-optical cluster state, we can write this
expression as
\begin{eqnarray}
\sum_k \left( \eta_k^2 \left\langle\hop{\delta}{k}^2\right\rangle
+ \sum_{l\neq k} \eta_k\eta_l\left\langle\hop{\delta}{k}\hop{\delta}{l}\right\rangle \right)
= \sum_k \left(3\eta_k^2+\sum_{l\neq k} 2\eta_k\eta_l\right)\times\frac{e^{-2r}}{4} \,,
\label{noise_gen_lo}
\end{eqnarray}
where we have applied \eref{miracle} in reaching the right hand
side. Since $\sum_k\eta_k=1$, it follows that
\begin{eqnarray}
\sum_{k,l} \eta_k\eta_l = \sum_k \left( \eta_k^2 + \sum_{l\neq k} \eta_k\eta_l \right) = 1 \,.
\end{eqnarray}
We can therefore reduce the expression \eref{noise_gen_lo} for the excess noise to the form
\begin{eqnarray}
\left( \sum_k\eta_k^2+2\right)\times\frac{e^{-2r}}{4} \,.
\label{noise_gen_lo-2}
\end{eqnarray}
It is now clear that minimization for the excess noise here is
completely identical to that for the case of canonical clusters (see
below \eref{q_gen_NR}). Namely, it corresponds to locating the point
over the $N$-dimensional hyperplane $\sum_k \eta_k=1$ that is
closest to the origin. Obviously, the result is the symmetric point
$\eta_k=1/N$ for all $k$, which means that the excess noise for
$\hop{q}{\sigma}$ in \eref{q_gen_NR_lo} can be minimized if it is
distributed equally over each of the multi-rails.

\section{\label{sec:1QND}Single-QND construction for multi-rail CV canonical clusters}
\begin{figure}
\includegraphics*[width=170mm]{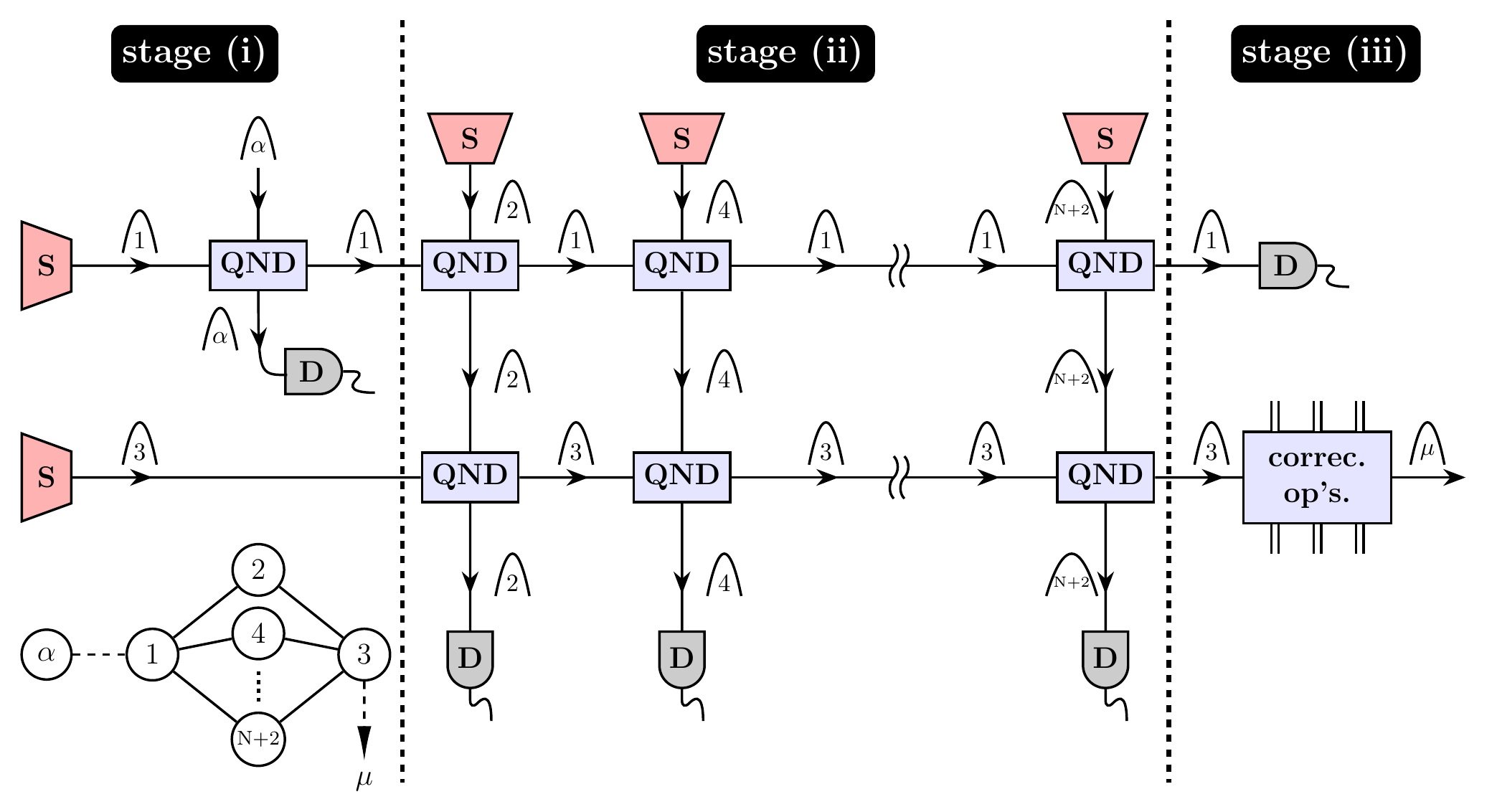}
\caption{A network of QND-gates for the single-mode teleportation through an $N$-rail
canonical cluster shown in the lower left corner of this figure. Here ``S" are single-mode squeezers
that generate momentum-squeezed vacuum modes, ``QND" are QND-gates that entangle the pairs of
intersecting modes, and ``D" the homodyne detectors for the qumodes.
Each qumode is denoted by a numbered pulse for the corresponding cluster node,
and similarly for the input mode $\alpha$ and the output mode $\mu$.
The teleportation is divided into three stages (see text), which are indicated by vertical dotted
lines. The gate ``correc.~op's" in stage (iii) represents corrective
operations for the output mode, where the double wires on the sides stand for signals
fed forward from measurement outcomes of other qumodes.
\label{fig:QND-net}}
\end{figure}

We discuss in this Appendix the fabrication of multi-rail CV
canonical cluster states in the time-encoding single-QND approach
\cite{Me10}. For simplicity, we will demonstrate with the
single-mode teleportation shown in Fig.~\ref{fig:MR_tel}(c), which
is redrawn in Fig.~\ref{fig:QND-net} with the nodes numbered in
accordance with their time sequence in the single-QND scheme (see
below). Let us start by considering the teleportation through an
extended network of QND-gates illustrated in Fig.~\ref{fig:QND-net},
where the teleportation is divided into three stages: (i) input
coupling, (ii) multi-rail construction, and (iii) output generation.
At stage (i), the input mode is coupled to qumode 1 through a
QND-gate. The input mode is then subject to a homodyne detection,
while qumode 1 is directed towards the next stage. At the same time,
in preparation for stage (ii), the momentum-squeezed mode 3 is also
generated. At stage (ii), qumodes 1 and 3 entangle with additional
momentum-squeezed qumodes $\{2, 4, \dots, (N+2)\}$ through sequence
of QND-gates to form the multi-rails. The ``mid-rail" modes $\{2, 4,
\dots, (N+2)\}$ in the multi-rails are measured subsequently in
preparation for the corrective operations over qumode 3 in stage
(iii), while modes 1 and 3 are led to the next stage. At stage
(iii), qumode 1 is measured, while qumode 3 is corrected in
accordance with measurement outcomes of all other modes to generate
the desired output mode $\mu$.

\begin{figure}
\includegraphics*[width=50mm]{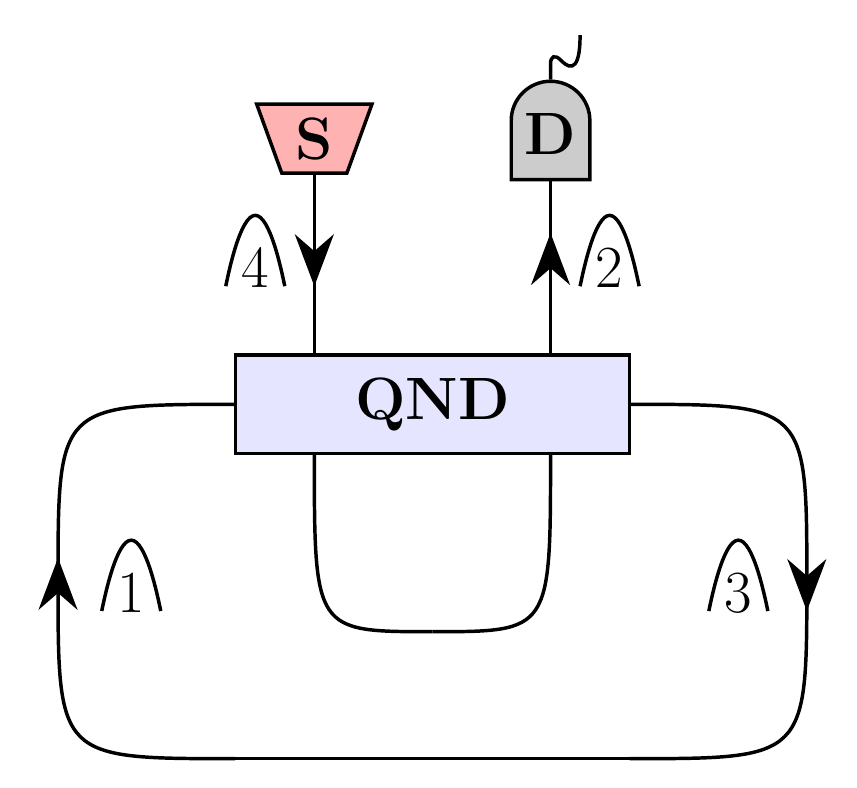}
\caption{A snapshot for the single-QND counterpart for stage (ii) of the multi-rail
teleportation depicted in Fig.~\ref{fig:QND-net}. Here the mid-rail modes
$\{2,4,\dots,(N+2)\}$ are produced sequentially from the squeezer at a fixed interval, so that they
can meet qumodes 1 and 3 at appropriate times at the QND gate.
Note that for the optical paths the path lengths illustrated here are not
in actual proportions and the loop for qumodes 1 and 3
is in fact part of a helical path; see text and also Fig.~\ref{fig:1QND-1}.
\label{fig:1QND-2}}
\end{figure}

\begin{figure}
\includegraphics*[width=170mm]{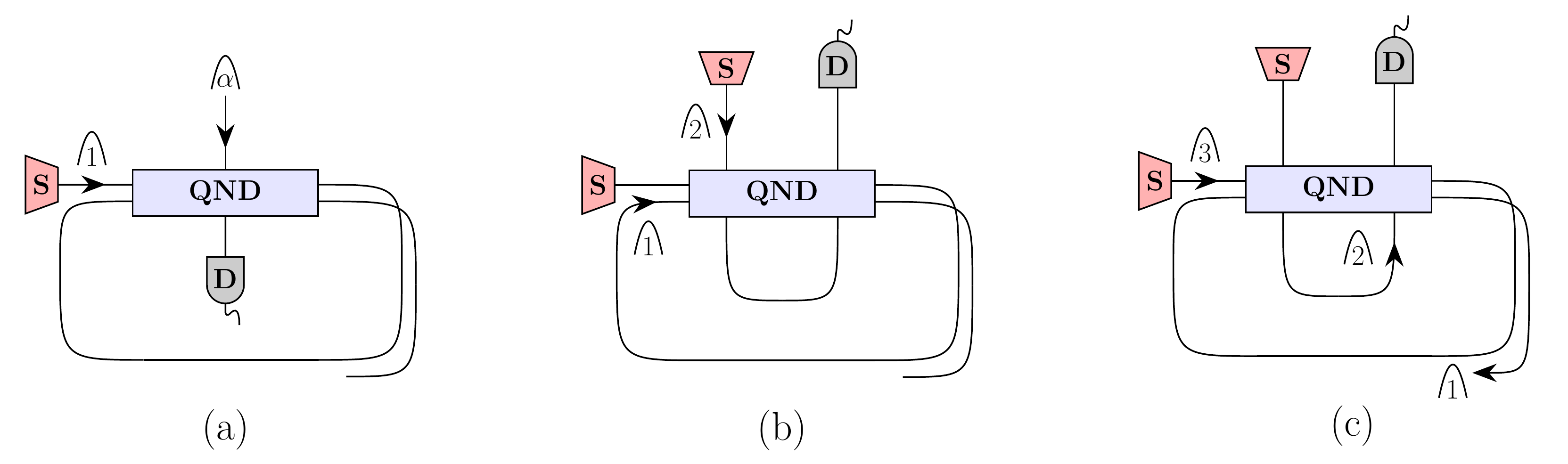}
\caption{Single-QND realization for the transition from stage (i) [panel (a)]
to the initial steps of stage (ii) [panels (b), (c)]
for the teleportation shown in Fig.~\ref{fig:QND-net}. For clarity, we show
only part of the helical path for modes 1 and 3.
\label{fig:1QND-1}}
\end{figure}

As is clear from Fig.~\ref{fig:QND-net}, stage (ii) of the
teleportation consists of $N$ repeated units that entangle modes 1
and 3 to the mid-rail modes $\{2, 4, \dots (N+2)\}$ via pairs of
QND-gates, and subsequently detect the mid-rail modes. One can thus
readily condense stage (ii) into the single-QND design shown in
Fig.~\ref{fig:1QND-2}. Namely, by cycling qumodes 1 and 3 repeatedly
through the QND-gate while directing the mid-rail modes along a
$U$-shape optical path that would meet qumodes 1 and 3 at
appropriate times, one would be able to implement the multi-rail
structure in the time domain utilizing a single QND-gate. Since
there are $N$ multi-rails in total, qumodes 1 and 3 must cycle $N$
times before entering stage (iii). Therefore, the optical path for
qumodes 1 and 3 can be fashioned into a helical one, so that the
loop in Fig.~\ref{fig:1QND-2} is just a single run of it. On the
other hand, due to the lack of symmetry in the setup at stage (i),
its transition into the highly symmetric stage (ii) poses the major
challenge to the single-QND implementation here. We propose to
overcome this difficulty through the design illustrated in
Fig.~\ref{fig:1QND-1}, as we shall now explain.

Firstly, we send in the input mode $\alpha$ towards the QND-gate for
coupling with qumode 1 at time $t=0$ and then direct it to a
homodyne detector [Fig.~\ref{fig:1QND-1}(a)]. Secondly, the
input-mode-encoded qumode 1 then proceeds along a helical path and
returns to the QND-gate for entanglement with qumode 2 at time
$t=T$. Qumode 2 then follows a $U$-shape optical path that will lead
it back to the QND-gate [Fig.~\ref{fig:1QND-1}(b)]. Thirdly, qumode
3 is then generated and propagates along the same optical path as
qumode 1, so that it would meet with qumode 2 at the QND-gate at
time $t=1.5\,T$. After this, qumode 2 is then detected, while qumode
3 proceeds along the helical path with a time lag $1.5\,T$ behind
qumode 1 [Fig.~\ref{fig:1QND-1}(c)]. We are now connected to the
fully implemented stage (ii) illustrated in Fig.~\ref{fig:1QND-2},
which corresponds roughly to a snapshot at $t\simeq 1.75\,T$.
Finally, in order to separate qumodes 1 and 3 over the same helical
path at stage (iii), one can prepare them with orthogonal
polarizations at the outset. The single-QND counterpart for stage
(iii) can then be achieved through a polarization beam-splitter
which would divert, for instance, qumode 1 to a homodyne detector,
while passing qumode 3 to the corrective gates. In this case, of
course, one must ensure that the QND-gate should function
impartially for both polarizations.

It is interesting to note that the scheme above can be generalized
easily for single-qumode teleportation involving a longer linear
cluster implemented with multi-rail structures. This can be done by
repeating the single-QND procedures above while skipping at stage
(iii) the corrective operations for qumode 3. That is, by taking the
{\em uncorrected} qumode 3 as the new input state $\alpha$ in
Fig.~\ref{fig:1QND-1}(a), one can furnish an additional multi-rail
teleportation down the linear cluster. Depending on the length of
the linear cluster, this iteration can be terminated via the stage
(iii) implementation above once the desired number of runs is
reached.

For two-mode operations, such as the CZ-gate teleportation of
Fig.~\ref{fig:MR_cz}(d), one could extend the preceding single-QND
design by either time multiplexing or polarization multiplexing
similar to what was proposed for two-dimensional square lattices in
the single-QND approach \cite{Me10}. For instance, if we modify the
node configurations in Fig.~\ref{fig:MR_cz}(d) slightly by
connecting the input modes to nodes 1 and $(2N+4)$, and output modes
to nodes $(N+2)$ and $(N+3)$, we would then have two identical,
independent single-qumode teleporting arms before reaching the
output stage. Therefore, the single-QND implementation here can be
achieved through two copies of the pulse sequence above
intercalating each other in time. At the output stage, however,
prior to subjecting qumodes $(N+2)$ and $(N+3)$ to the respective
corrective operations, one must direct them back to the QND-gate for
entanglement. Since these two modes are separated in time, this can
be achieved with the help of an active beam-divertor.



\begin{thebibliography}{99}
\bibitem{note2}
By ``inline squeezing", we mean squeezing operations over non-vacuum input states, while
``offline squeezing" refers to those with vacuum inputs.
\bibitem{note3}
Strictly speaking, the quadrature manipulation for the case of Fig.~\ref{fig:L4-1}
belongs also not to the trivial case (a), since the corrective operations
\eref{corr_L4-1} involve Fourier transformations over output cluster nodes. Thus,
the quadrature manipulation in this case belongs to category (c). Nevertheless, the
Fourier transformations $\hop{F}{k}^2$ in \eref{corr_L4-1} merely induce sign changes
in the quadrature operators for nodes $k=2,3$, the remaining quadrature manipulations
there are relatively trivial in comparison with other cases considered in this paper.
\bibitem{note35}
Note that the number 1/4 is dependent on our choice of setting
$\hbar=1/2$ in writing \eref{qp_def}. Another choice with $\hbar=2$
that is often adopted in the related literatures would have this
number being 1, instead; see \cite{We12}.
\bibitem{note4}
For the case of Fourier-transformed input modes, the input-coupling
equation \eref{in_coup2} for canonical cluster states remain to have
$\hat{q}''$ entries that won't be helpful for quadrature
manipulations, despite the Fourier transformations. This is because
these entries involve only either the input mode $\rho$ or the
cluster mode $k$, while for quadrature manipulations we need entries
of the input-coupling equation that would mix the input and the
cluster modes.
\end{thebibliography}
\end{document}